\documentclass{article}
\pdfoutput=1

\usepackage{arxiv}

\usepackage[utf8]{inputenc} % allow utf-8 input
\usepackage[T1]{fontenc}    % use 8-bit T1 fonts
\usepackage{hyperref}       % hyperlinks
\usepackage{url}            % simple URL typesetting
\usepackage{booktabs}       % professional-quality tables
\usepackage{amsfonts}       % blackboard math symbols
\usepackage{nicefrac}       % compact symbols for 1/2, etc.
\usepackage{microtype}      % microtypography
\usepackage{lipsum}		% Can be removed after putting your text content
\usepackage{graphicx}
\usepackage{natbib}
\usepackage{doi}

\usepackage{amsmath}
\usepackage{algorithm}
\usepackage{algpseudocode}
\usepackage{multirow}
\usepackage[percent]{overpic}

\title{ECG-based estimation of respiratory modulation of \\
AV nodal conduction during atrial fibrillation}

\author{ \href{https://orcid.org/0000-0002-1028-4276}{\includegraphics[scale=0.06]{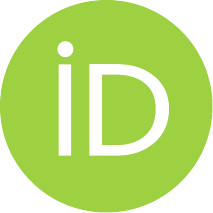}\hspace{1mm}Felix Plappert}\\
	Department of Biomedical Engineering\\
	Lund University, \\
    Lund, Sweden \\
	\texttt{felix.plappert@bme.lth.edu} \\
	\And
	\href{https://orcid.org/0000-0002-8618-9152}{\includegraphics[scale=0.06]{orcid.pdf}\hspace{1mm}Gunnar Engstr\"om} \\
	Department of Clinical Sciences, \\
    Cardiovascular Research -- Epidemiology\\
	Malm\"o, Sweden\\
    \And
	\href{https://orcid.org/0000-0002-5592-8717}{\includegraphics[scale=0.06]{orcid.pdf}\hspace{1mm}Pyotr G.~Platonov} \\
	Department of Cardiology, Clinical Sciences, \\
    Lund University, \\
    Lund, Sweden\\
    \And
	\href{https://orcid.org/0000-0002-2519-736X}{\includegraphics[scale=0.06]{orcid.pdf}\hspace{1mm}Mikael Wallman} \\
	Fraunhofer-Chalmers Centre, \\
    Department of Systems and Data Analysis\\
    Gothenburg, Sweden\\
    \And
	\href{https://orcid.org/0000-0002-1956-6333}{\includegraphics[scale=0.06]{orcid.pdf}\hspace{1mm}Frida Sandberg} \\
	Department of Biomedical Engineering\\
	Lund University,\\
    Lund, Sweden \\
}

\renewcommand{\shorttitle}{Respiratory modulation estimation during AF}

\hypersetup{
pdftitle={ECG-based estimation of respiratory modulation of AV nodal conduction during atrial fibrillation},
pdfsubject={eess.SP, q-bio.TO},
pdfauthor={Felix Plappert, Gunnar Engstr\"om, Pyotr G.~Platonov, Mikael Wallman, Frida Sandberg},
pdfkeywords={atrial fibrillation, atrioventricular node, autonomic tone, respiratory modulation, convolutional neural network, deep breathing test, network model, ECG},
}

\begin{document}
\maketitle

\begin{abstract}
Information about autonomic nervous system (ANS) activity may be valuable for personalized atrial fibrillation (AF) treatment but is not easily accessible from the ECG.
In this study, we propose a new approach for ECG-based assessment of respiratory modulation in AV nodal refractory period and conduction delay.
A 1-dimensional convolutional neural network (1D-CNN) was trained to estimate respiratory modulation of AV nodal conduction properties from 1-minute segments of RR series, respiration signals, and atrial fibrillatory rates (AFR) using synthetic data that replicates clinical ECG-derived data.
The synthetic data were generated using a network model of the AV node and 4 million unique model parameter sets.
The 1D-CNN was then used to analyze respiratory modulation in clinical deep breathing test data of 28 patients in AF, where a ECG-derived respiration signal was extracted using a novel approach based on periodic component analysis.
We demonstrated using synthetic data that the 1D-CNN can predict the respiratory modulation from RR series alone ($\rho=0.805$) and that the addition of either respiration signal ($\rho=0.830$), AFR ($\rho=0.837$), or both ($\rho=0.855$) improves the prediction.
Results from analysis of clinical ECG data of 20 patients with sufficient signal quality suggest that respiratory modulation decreased in response to deep breathing for five patients, increased for five patients, and remained similar for ten patients, indicating a large inter-patient variability.
\end{abstract}

% keywords can be removed

\section{Introduction}
Atrial fibrillation (AF) is the most common supraventricular tachyarrhythmia (\cite{Hindricks2020}).
Characteristic for AF is an uncoordinated atrial electrical activation that results in an increased and irregular ventricular activity.
Atrial fibrillation poses a significant burden to patients, physicians, and healthcare systems globally, and is associated with substantial morbidity and mortality.
Two main strategies of AF treatments are rate control and rhythm control.
Rate control is one of the cornerstones of AF management, but the effect of individual rate-control drugs is difficult to predict in advance.
Therefore, to personalize the AF treatment and reduce the AF burden on the healthcare system, it is essential to better understand the complex mechanisms of AF.

The autonomic nervous system (ANS) contributes to the initiation and maintenance of AF (\cite{Shen2014}), where a predominance in both the sympathetic and parasympathetic modulation has been observed to initiate episodes of paroxysmal atrial fibrillation (\cite{Lombardi2004}).
Thus, differences in ANS activity among patients may explain part of the inter-patient variability in AF treatment response.
If the ANS activity of individual AF patients could be quantified, it would allow for a better understanding of the role that the ANS plays in patient-specific AF pathophysiology.
In normal sinus rhythm (NSR), heart rate variability (HRV) is commonly used to obtain information about the autonomic tone and respiratory modulation (\cite{Sassi2015,Shaffer2017}).
From the ventricular rhythm, HRV assesses the function of the sinoatrial node, which is densely innervated by the ANS, and allows conclusions about the autonomic tone 
(\cite{Shen2014,George2017}).
For instance, deep breathing known to elicit a parasympathetic response has a pronounced effect on the high-frequency component of the HRV during sinus rhythm (\cite{Bernardi2001}).
In AF, however, the interpretation of HRV differs because the ventricular rhythm is instead a product of the atrial electrical activity and subsequent AV nodal modulation.
Since the AV node is also densely innervated by the ANS, the behavior of the AV node during AF may give valuable information about the autonomic tone.
However, the relation between the heart rate characteristics and the AV nodal conduction properties during AF is non-trivial.
For this reason, there is a demand for a methodology that can quantify the effect of the ANS in individual AF patients, and a model-based analysis of the AV nodal properties may be a viable approach.

The AV node is characterized by its dual-pathway physiology allowing for parallel conduction of impulses where the two pathways have different electrophysiological properties (\cite{George2017}).
The fast pathway (FP) exhibits a shorter conduction delay and longer refractory period compared to the slow pathway (SP) (\cite{George2017}).
The AV nodal refractory period and conduction delay are history dependent and are influenced by the previous activity of conducted and blocked impulses (\cite{George2017, Billette2019}).
There have been several AV node models proposed that describe different characteristics of the AV nodal structure and electrophysiology (\cite{Cohen1983,Rashidi2005,Mangin2005,Lian2006,Climent2011,Mase2015,Henriksson2016,Inada2017,Wallman2018,Karlsson2021}), but our previously proposed model (\cite{Plappert2022}) is the first to address ANS-induced changes to the AV nodal refractory period and conduction delay.
We showed that ANS-induced changes during tilt could be better replicated when scaling the refractory period and conduction delay with a constant factor.
However, the variation in sympathetic and parasympathetic activity is essential for the autonomic function and the AV node model should account for time-varying changes in the modulation of AV nodal refractory period and conduction delay.

Machine learning is vibrant in the field of cardiac electrophysiology with a rapidly growing number of applications (\cite{Trayanova2021}).
However, one main challenge is the acquirement of large amounts of data for proper training and validation.
In recent years, a few studies have been performed in which synthetic data has been generated for the training of neural networks which are then used on clinical data.
For example, synthetic images were generated to train neural networks to track cardiac motion and calculate cardiac strain (\cite{Loecher2021}), to estimate tensors from free-breathing cardiac diffusion tensor imaging (\cite{Weine2022}), and to predict end-diastole volume, end-systole volume and ejection fraction (\cite{Gheorghita2022}).
Furthermore, synthetic photoplethysmography (PPG) signals were generated to detect bradycardia and tachycardia (\cite{Solosenko2022}), and synthetic electrocardiogram (ECG) signals were generated to detect r-waves during different physical activities and atrial fibrillation (\cite{Kaisti2023}), and to predict the ventricular origin in outflow tract ventricular arrhythmias (\cite{Doste2022}).

This study aims to develop and evaluate a method to extract respiratory modulation in the AV node conduction properties from ECG data in AF.
We present a novel approach to extract respiration signals from several ECG leads based on the periodic component analysis (\cite{Sameni2008}).
In addition, we present a novel extension to our previously proposed AV node network model accounting for respiratory modulation of AV nodal refractory period and conduction delay.
Furthermore, we predict the magnitude of respiratory modulation using a 1-dimensional convolutional neural network that was trained on synthetic 1-minute segments of RR series, respiration signals, and average atrial fibrillatory rate which replicate clinical data.
The trained network was used to analyze data from 28 AF patients performing a deep breathing task.

\section{Materials and methods}
First, the clinical deep breathing test data from patients in atrial fibrillation is described in Sec. \ref{sec:clinicaldata}.
In Sec. \ref{sec:pica}, the extraction of RR series and atrial fibrillatory rate (AFR) from ECG are described.
Moreover, Sec. \ref{sec:pica} covers the extraction of ECG-derived respiration (EDR) signals using a novel approach based on periodic component analysis.
A description of the extended AV network model accounting for respiratory modulation is given in Sec. \ref{sec:simulateddata}, as well as a description of how the simulated datasets are generated.
In Sec. \ref{sec:recurrentneuralnetwork}, the architecture of a 1-dimensional convolutional neural network (1D-CNN) that is used to predict the magnitude of respiratory modulation from ECG recordings is described together with the training and testing of the neural network.
Finally, the CNN is used to predict the respiratory modulation in clinical data and the predictions are analyzed.

\subsection{ECG data}
\label{sec:clinicaldata}
The dataset of the clinical deep breathing test consisted of 12-lead ECG recordings with a sampling rate of 500\,Hz from individuals with AF participating in the SCAPIS study (\cite{Bergstrom2015}).
The participants in the SCAPIS study were from the Swedish general population aged 50-64 years.
A subset of the SCAPIS cohort (5136 participants) performed a deep breathing test (\cite{Engstrom2022}).
Of this subset, 28 participants with complete data were in AF at the time of recording (\cite{Abdollahpur2022}).
The clinical characteristics of that subset are listed in Table \ref{tab:1}.
\begin{table}
\caption{\label{tab:1} Clinical characteristics of study population.}
\vspace{4 mm}
\centerline{\begin{tabular}{ll} \hline\hline
 & Number  \\ \hline
Age & 60.1\,$\pm$\,4.0 [50.1-64.9] \\
Men & 23 (82\%)\\
BMI & 31.8\,$\pm$\,7.2 [18.8-50.8] \\
Systolic BP & 124\,$\pm$\,23 [90-188] \\
Diastolic BP & 79.9\,$\pm$11 [61-104] \\
Hypertension$^{\ast}$ & 17 (61\%) \\
Diabetes & 2 (7\%) \\
Never smokers & 9 (32\%) \\
Heart failure & 2 (7\%) \\
Previous AMI or angina & 2 (7\%) \\
\hline \\
Treatment & \\
Beta blocker & 15 (54\%) \\
Ca-antagonist & 6 (21\%) \\
Antiarrhythmic drug & 4 (14\%) \\
\hline\hline\\
\multicolumn{2}{l}{${}^{\ast}$\,$\geq$\,140/90\,mmHg or treatment for hypertension}\\
\multicolumn{2}{l}{Values are given in the following formats: number, mean\,$\pm$\,SD, [range]; BP, blood pressure.}\\
\end{tabular}}
\end{table}
The deep breathing test started with the participants resting in a supine position while breathing normally for five minutes.
Then for the deep breathing, the participants were guided by a nurse to inhale for five seconds and exhale for five seconds, for a total of six breathing cycles corresponding to 1 minute.

\subsection{ECG data processing}
\label{sec:pica}
\subsubsection{Extraction of RR series}
\label{sec:ecgprocessing}
ECG preprocessing and QRS complex detection were performed using the CardioLund ECG parser (www.cardiolund.com).
The CardioLund ECG parser classified QRS complexes based on their QRS morphology.
Only QRS complexes with dominant QRS morphology were considered in the computation of the RR series.

The RR series were computed from intervals between R-peaks taken from consecutive QRS complexes with dominant QRS morphology, and the time of each RR interval was set to the time of the first R-peak in each interval.
The resulting non-uniformly sampled RR series were interpolated to a uniform sampling rate of 4\,Hz using piecewise cubic Hermite polynomials as implemented in MATLAB ('pchip', version R2023a, RRID:SCR\_001622).

\subsubsection{Estimation of atrial fibrillatory rate}
\label{sec:AFRest}
The AFR was used to obtain information about the atrial arrival process.
Briefly, the estimation of the AFR involved the extraction of an f-wave signal by means of spatiotemporal QRST-cancellation (\cite{Stridh2001}) and estimation of an f-wave frequency trend by fitting two complex exponential functions to the extracted f-wave signal from ECG lead V1 as proposed in (\cite{Henriksson2018}).
The two exponential functions were characterized by a fundamental frequency $f$ and its second harmonic, respectively; $f$ was fitted within the range $f_{max}^{Welch}\pm 1.5$\,Hz, where $f_{max}^{Welch}$ denotes the maximum of the Welch periodogram of ECG lead V1 in the range $4-12$\,Hz.
The results for the deep breathing data have been previously presented in (\cite{Abdollahpur2022}).
The estimated AFR signal has a sampling rate of 50\,Hz.

\subsubsection{Extraction of lead-specific EDR signals}
\label{sec:extractionleadspec}
All steps of the extraction algorithm that are described in the following were applied to \mbox{1-minute} segments of the lead-specific EDR signals taken from a \mbox{1-minute} running window.
The lead-specific EDR signals were computed with the slope range method (\cite{Kontaxis2020}) for the eight ECG leads V1-V6, I, and II. 
Only eight out of 12 ECG leads were used, because the information in the leads III, aVF, aVL, and aVR can also be derived from lead I and II.
The slope range method uses the peak-to-peak difference in the first derivative of the QRS complex to quantify the variations in the QRS morphology that are assumed to reflect the respiratory activity and are caused for example by periodic changes in electrode positions relative to the heart.

Only QRS complexes with dominant QRS morphology (cf. Sec. \ref{sec:ecgprocessing}) were considered when applying the slope-range method.
Further, a QRS complex was excluded as an outlier from analysis if the slope range value of any of the leads was outside the mean\,$\pm$\,3\,std of the slope range values of that lead.
The lead-specific non-uniformly sampled EDR signals were interpolated to a uniform sampling rate of 4\,Hz using the modified Akima algorithm as implemented in MATLAB ('makima', version R2023a, RRID:SCR\_001622).
A matrix containing the resampled lead-specific EDR signals $\mathbf{X}'=[\mathbf{x}'_1,...,\mathbf{x}'_8]^{\mkern-1.5mu\mathsf{T}}$ of dimension $8\!\times\! N$ was constructed, where $N=240$ corresponds to the length of the \mbox{1-minute} segment.
To remove baseline-wander in $\mathbf{X}'$, a Butterworth highpass filter of order 4 with a cut-off frequency of 0.08\,Hz was applied separately for each lead $\mathbf{x}'$.
The filtered $\mathbf{X}'$ was normalized to zero-mean and signals shorter than 1 minute were zero-padded to create $\mathbf{X}$ containing \mbox{1-minute} segments.
A set $\mathcal{S}_{seg}$ was created containing all $\mathbf{X}_i$, where $i=1,...,I$ denotes all $I$ possible choices of \mbox{1-minute} segments of the lead-specific EDR signals from one recording.

\subsubsection{Extraction of joint-lead EDR signals}
\label{sec:extractionjointleadEDR}
The joint-lead EDR signal was extracted from $\mathbf{X}$ using a modified version of the periodic component analysis ($\pi$CA) (\cite{Sameni2008}), summarized in Alg. \ref{alg:PiCA}.
\begin{algorithm}
\caption{Extraction of joint-lead EDR signals}\label{alg:PiCA}
\begin{algorithmic}
\For{all $\mathbf{X}_i$ in $\mathcal{S}_{seg}$}
    \State $\mathbf{X}_i$ is whitened according to Eq. \ref{eq:whitening} to obtain $\mathbf{Z}_i$
    \For{all $\tau_j\in[10,40]$}
    \State obtain $\mathbf{w}_j$ by solving the generalized eigenvalue problem of matrix pair $(\overline{\mathbf{C}}_{\mathbf{z}}(\tau_j),\overline{\mathbf{C}}_{\mathbf{z}}(0))$
    \State compute $\epsilon(\mathbf{w}_j,\tau_j,\mathbf{Z}_i)$ according to Eq. \ref{eq:epsilon}
    \EndFor
\EndFor
\State compute $\tau^{\ast}=\min_{\tau_{j}}\left(\sum_{\mathcal{S}_{seg}} \epsilon(\mathbf{w}_j,\tau_j,\mathbf{Z}_i)\right)$
\For{all $\mathbf{Z}_i$ in $\mathcal{S}_{seg}$}
\State $\mathcal{S}_{\tau}=\emptyset$
\For{all $\tau_j\in[10,40]$}
\If{$\epsilon(\mathbf{w}_j,\tau_j,\mathbf{Z}_i)\leq\epsilon(\mathbf{w}_j,\tau_{j-1},\mathbf{Z}_i) \lor \tau_j==10$}
\If{$\epsilon(\mathbf{w}_j,\tau_j,\mathbf{Z}_i)\leq\epsilon(\mathbf{w}_j,\tau_{j+1},\mathbf{Z}_i) \lor \tau_j==40$}
\State add $\tau_{j}$ to $\mathcal{S}_{\tau}$ 
\EndIf
\EndIf
\EndFor
\State set $\tau_{resp}$ as value in $\mathcal{S}_{\tau}$ closest to $\tau^{\ast}$
\State obtain $\mathbf{w}_{resp}$ by solving the generalized eigenvalue problem of matrix pair $(\overline{\mathbf{C}}_{\mathbf{z}}(\tau_{resp}),\overline{\mathbf{C}}_{\mathbf{z}}(0))$
\State $\mathbf{s}^{\ast}_i = \mathbf{w}_{resp}^{\mkern-1.5mu\mathsf{T}}\mathbf{Z}_i\cdot \textrm{sign}\left(\sum \mathbf{w}_{resp}\right)$
\State $f_{resp,i} = f_s/\tau_{resp}$
\EndFor
\end{algorithmic}
\end{algorithm}
%The original formulation of $\pi$CA was used to analyze the \mbox{1-minute} segments of the eight resampled lead-specific EDR signals $\mathbf{X}(n)$.
The matrix $\mathbf{X}$ was whitened for its elements to be uncorrelated and to have unit variance.
The whitened lead-specific EDR signals $\mathbf{Z}$ were computed as
\begin{equation}
\label{eq:whitening}
    \mathbf{Z} = \mathbf{D}^{-1/2}\mathbf{E}^{\mkern-1.5mu\mathsf{T}}\mathbf{X},
\end{equation}
where $\mathbf{D}$ is the diagonal matrix of eigenvalues of the covariance matrix $\mathbf{C}_{\mathbf{X}}=E\{\mathbf{X}\mathbf{X}^{\mkern-1.5mu\mathsf{T}}\}$, and the columns of the matrix $\mathbf{E}$ are the unit-norm eigenvectors of $\mathbf{C}_{\mathbf{X}}$.

The outputs of the $\pi$CA are a joint-lead EDR signal $\mathbf{s}$ of dimension $1\!\times\! N$ and its corresponding lag $\tau$.
The assumption of the $\pi$CA is that $\mathbf{s}=\mathbf{w}^{\mkern-1.5mu\mathsf{T}}\mathbf{Z}$ is a linear mixture of the whitened lead-specific EDR signals.
The aim is to find a solution for $\mathbf{s}$ with a maximal periodic structure.
The periodic structure of $\mathbf{s}$ is characterized by $\epsilon(\mathbf{w},\tau,\mathbf{Z})$, which quantifies non-periodicity (\cite{Sameni2008}) and is defined as
\begin{equation}
\label{eq:epsilon}
    \epsilon(\mathbf{w},\tau,\mathbf{Z}) = \dfrac{\sum_{n}|s(n+\tau)-s(n)|^2}{\sum_{n}|s(n)|^2} = 2\left[1-\dfrac{\mathbf{w}^{\mkern-1.5mu\mathsf{T}}\overline{\mathbf{C}}_{\mathbf{z}}(\tau)\mathbf{w}}{\mathbf{w}^{\mkern-1.5mu\mathsf{T}}\overline{\mathbf{C}}_{\mathbf{z}}(0)\mathbf{w}}\right],
\end{equation}
where $s(n)$ is the $n$:th element of $\mathbf{s}$.
We solved the generalized eigenvalue problem (GEP) of the lag-dependent matrix pair $(\overline{\mathbf{C}}_{\mathbf{z}}(\tau),\overline{\mathbf{C}}_{\mathbf{z}}(0))$ to obtain a full matrix $\mathbf{V}$ whose columns correspond to the right eigenvectors and a diagonal matrix $\mathbf{U}$ of generalized eigenvalues so that $\overline{\mathbf{C}}_{\mathbf{z}}(\tau)\mathbf{V}=\overline{\mathbf{C}}_{\mathbf{z}}(0)\mathbf{V}\mathbf{U}$ (\cite{Sameni2008}).
Here, \mbox{$\overline{\mathbf{C}}_{\mathbf{z}}(\tau)=1/4[\mathbf{C}_{\mathbf{z}}(\tau)+(\mathbf{C}_{\mathbf{z}}(\tau))^{\mkern-1.5mu\mathsf{T}}+\mathbf{C}_{\mathbf{z}}(-\tau)+(\mathbf{C}_{\mathbf{z}}(-\tau))^{\mkern-1.5mu\mathsf{T}}]$} for some lag $\tau$ is a modified lagged covariance matrix, which is always symmetric, unlike the time lagged covariance matrix \mbox{$\mathbf{C}_{\mathbf{z}}(\tau)=E_n\{\mathbf{z}(n)\mathbf{z}(n-\tau)^{\mkern-1.5mu\mathsf{T}}\}$}, where $\mathbf{z}(n)$ is the $n$:th column of $\mathbf{Z}$ and $E_n\{\cdot\}$ indicates averaging over $n$.
The weight vector \mbox{$\mathbf{w}=[w_1,...,w_8]^{\mkern-1.5mu\mathsf{T}}$} that minimizes $\epsilon(\mathbf{w},\tau,\mathbf{Z})$ is obtained as the first column of $\mathbf{V}$ (\cite{Sameni2008}).
In the present study, $\epsilon(\mathbf{w},\tau,\mathbf{Z})$ is also used to quantify signal quality, where a lower value of $\epsilon(\mathbf{w},\tau,\mathbf{Z})$ corresponds to a more periodic signal assumed to have a higher SNR.

As $\tau$ is unknown, $\epsilon(\mathbf{w},\tau,\mathbf{Z})$ was minimized for all integer values of $\tau$ between 10 and 40, corresponding to respiration rates between 0.1 and 0.4\,Hz.
%According to the assumption of the $\pi$CA, the $\tau$ that minimizes $\epsilon(\mathbf{w},\tau,\mathbf{Z})$ corresponds to the respiration frequency, but this is not always the case in data with lower signal quality.
To improve the robustness of the $\pi$CA for signals with low quality, a $\tau^{\ast}$ was determined in an intermediate step that corresponds to a global minimum of $\epsilon(\mathbf{w},\tau,\mathbf{Z})$ over all \mbox{1-minute} segments in $\mathcal{S}_{seg}$.
It was assumed that there were no significant transient changes in respiration frequency in the clinical data and we determined two different $\tau^{\ast}$ for each subject; one for normal breathing and one for deep breathing.
Then, for each \mbox{1-minute} segment separately, a $\tau_{resp}$ was estimated as the local minimum of $\epsilon(\mathbf{w},\tau,\mathbf{Z})$ closest to $\tau^{\ast}$.
The respiration frequency estimate $\hat{f}_{resp}=f_{s}/\hat{\tau}_{resp}$ results from the estimate $\hat{\tau}_{resp}$ and the sampling rate $f_s=4$\,Hz and is in the range $\hat{f}_{resp}\in[0.1,0.4]$Hz corresponding to the limits set by $\tau$.
Finally, the weight vector $\mathbf{w}_{resp}$ for the respiration signal extraction was obtained by solving the GEP of the matrix pair $(\overline{\mathbf{C}}_{\mathbf{z}}(\tau_{resp}),\overline{\mathbf{C}}_{\mathbf{z}}(0))$.
The extracted $\mathbf{s}=\mathbf{w}_{resp}^{\mkern-1.5mu\mathsf{T}}\mathbf{Z}$ was normalized to unit variance.
An ambiguity of $\pi$CA is that the sign of $\mathbf{s}$ is undetermined.
The sign of the joint-lead EDR signal was selected as $\mathbf{s}^{\ast} = \mathbf{s}\cdot \textrm{sign}\left(\sum \mathbf{w}_{resp}\right)$, where $\sum \mathbf{w}_{resp}$ denotes the sum of the elements in the vector $\mathbf{w}_{resp}$.
This was done under the assumption that all lead-specific EDR signals are in phase.

\subsubsection{Estimates from clinical data}
\label{sec:PostprocessingRRseriesEDRsignals}
The joint-lead EDR signal extraction from Sec. \ref{sec:extractionjointleadEDR} was applied to all 1-minute segments $\mathbf{X}$ in $\mathcal{S}_{seg}$ for each patient and recording.
Segments $\mathbf{X}$ were excluded from further analysis if they do not satisfy the following three criteria, for which a valid QRS complex has a dominant QRS morphology and is not classified as outlier based on its slope range values: $i)$ the maximum distance between valid QRS complexes is 2 seconds; $ii)$ the minimum number of valid QRS complexes in a \mbox{1-minute} segment is 48; $iii)$ the minimum number of valid QRS complexes in a \mbox{1-minute} segment is 80\% of the normal-to-normal average heart rate of the \mbox{1-minute} segment. 
After exclusion, several sets of non-overlapping 1-minute segments could be created from the remaining $\mathbf{X}$.
Out of these, the set $\mathcal{S}_{seg}^{\ast}$ that resulted in the smallest sum of $\epsilon(\mathbf{w}_{resp},\tau_{resp},\mathbf{Z})$ was chosen, and used to produce joint-lead EDR signals $\mathcal{X}_{Resp}^{Clin}$ of dimension $1\!\times\! N$ as described in Sec. \ref{sec:extractionjointleadEDR}.
In addition, the corresponding \mbox{1-minute} RR series $\mathcal{X}_{RR}^{Clin}$ of dimension $1\!\times\! N$ was extracted from the RR series obtained in Section \ref{sec:ecgprocessing}.
We estimated the mean arrival rate of atrial impulses to the AV node $\hat{\mu}$ as $1000/\overline{AFR}$, where $\overline{AFR}$ is the average AFR-trend within each of the selected \mbox{1-minute} windows as described in Sec. \ref{sec:AFRest}.
To match the dimensions of $\mathcal{X}_{RR}^{Clin}$ and $\mathcal{X}_{Resp}^{Clin}$, $\hat{\mu}$ was then repeated $N$ times to produce $\mathcal{X}_{AFR}^{Clin}$ of dimension $1\!\times\! N$.
From the clinical data, a maximum of five non-overlapping \mbox{1-minute} long segments in normal breathing and one segment in deep breathing was obtained for $\mathcal{X}_{RR}^{Clin}$, $\mathcal{X}_{Resp}^{Clin}$ and $\mathcal{X}_{AFR}^{Clin}$.

\subsection{Simulated Data}
\label{sec:simulateddata}
\subsubsection{Network model of the human atrioventricular node}
\label{sec:networkmodel}
\begin{figure}
    \centering
    \includegraphics[width=\textwidth]{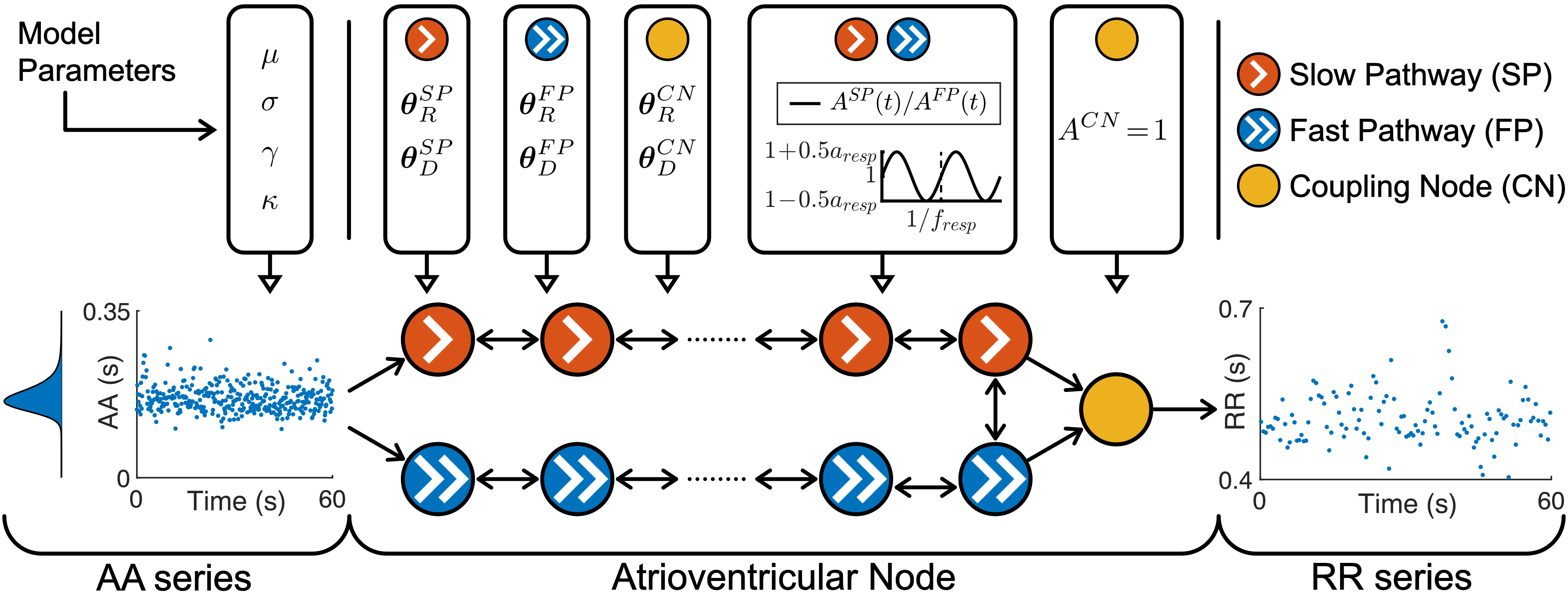}
    \caption{A schematic representation of the AV node model. Retrograde conduction was possible within the AV node model. For simplicity, only a subset of the ten nodes in each pathway is shown. Note that the atrioventricular node used different  $\boldsymbol{\theta}^{P}_{R}$ and $\boldsymbol{\theta}^{P}_{D}$ for the three different pathways, the same time-varying $A^{P}(t)$ for SP and FP and a constant $A^{CN}=1$ for CN.}
    \label{fig:1}
\end{figure}

The atrioventricular node is modeled by a network of 21 nodes (cf. Fig. \ref{fig:1}).
The presented AV node model was initially proposed in (\cite{Wallman2018}), updated with minor modifications in (\cite{Karlsson2021}), and extended using constant scaling factors $A_R$ and $A_D$ for the refractory period and conduction delay to account for the effect of changes in the autonomic tone in (\cite{Plappert2022}).
The slow pathway (SP) and fast pathway (FP) are described by two chains of 10 nodes each, which are only connected at their last nodes.
Impulses enter the AV node model simultaneously at the first node of each pathway.
Within the pathways and between their last nodes, the impulses are conducted bidirectionally to allow for retrograde conduction.
The last nodes of the two pathways are connected to an additional coupling node (CN), through which the impulses leave the model.

Each node represents a section of the AV node and is characterized by an individual refractory period $R^{P}\!(\Delta t_{k},A^{P}(t),\boldsymbol{\theta}^{P}_{R})$ and conduction delay $D^{P}\!(\Delta t_{k},A^{P}(t),\boldsymbol{\theta}^{P}_{D})$ defined as
\begin{align}
    \label{eq:Rmod}
    R^{P}\!\left(\Delta t_{k},A^{P}(t),\boldsymbol{\theta}^{P}_{R}\right) &= A^{P}(t) \left(R_{min}^{P} + \Delta R^{P}\left(1-e^{-\Delta t_{k}/\tau_R^{P}}\right)\right) \\
    \label{eq:Dmod}
    D^{P}\!\left(\Delta t_{k},A^{P}(t),\boldsymbol{\theta}^{P}_{D}\right) &= A^{P}(t)\left(D_{min}^{P} + \Delta D^{P}e^{-\Delta t_{k}/\tau_D^{P}}\right)
\end{align}
where $P\in\{SP,FP,CN\}$ denotes the pathway.
The refractory period and conduction delay are defined by fixed model parameters for the refractory period $\boldsymbol{\theta}^{P}_{R}$ and conduction delay $\boldsymbol{\theta}^{P}_{D}$ as well as model states for the diastolic interval $\Delta t_k$ and respiratory modulation $A^P(t)$.
Each pathway has a separate set of fixed model parameters for the refractory period $\boldsymbol{\theta}^{P}_{R}=[R_{min}^{P}$, $\Delta R^{P}$, $\tau_{R}^{P}]$ and conduction delay $\boldsymbol{\theta}^{P}_{D}=[D_{min}^{P}$, $\Delta D^{P}$, $\tau_{D}^{P}]$, where $R_{min}^{P}$ is the minimum refractory period, $\Delta R^{P}$ is the maximum prolongation of the refractory period, $\tau_{R}^{P}$ is a time constant, $D_{min}^{P}$ is the minimum conduction delay, $\Delta D^{P}$ is the maximum prolongation of the conduction delay and $\tau_{D}^{P}$ is a time constant.
For clarity, the notation of $R^{P}\!(\cdot,A^{P}(t),\cdot)$ and $R^{P}\!(\cdot,A^{P}(t),\cdot)$ are specified with dots when the replaced parameters or model states are currently not discussed.

The scaling factor $A^{P}(t)$ accounts for the effect of changes in autonomic tone on the refractory period $R^{P}\!\left(\cdot,A^{P}(t),\cdot\right)$ and the conduction delay $D^{P}\!\left(\cdot,A^{P}(t),\cdot\right)$.
The time-varying scaling factor $A^{P}(t)$ is common between the SP and FP, defined as
\begin{equation}
    \label{eq:ard}
    A^{SP}(t) = A^{FP}(t) = 1 + \dfrac{a_{resp}}{2}\,\textrm{sin}(2\pi t f_{resp}),
\end{equation}
with a constant respiratory frequency $f_{resp}$ and peak-to-peak amplitude $a_{resp}$.
The scaling factor of the refractory period and conduction delay of the CN is described by $A^{CN} = 1$ and not modulated by respiration.

The electrical excitation propagation through the AV node is modeled as a series of impulses that can either be conducted or blocked by a node.  
An impulse is conducted to all adjacent nodes, if the interval $\Delta t_{k}$ between the $k$:th impulse arrival time $t_{k}$ and the end of the ($k$--$1$):th refractory period, computed as 
\begin{equation}
    \label{eq:deltati}
    \Delta t_{k} = t_{k}-t_{k-1}-R^{P}\!\left(\Delta t_{k},\cdot,\cdot\right)
\end{equation} is positive.
Then, the time delay between the arrival of an impulse at a node and its transmission to all adjacent nodes is given by the conduction delay $D^{P}\!\left(\Delta t_{k},\cdot,\cdot\right)$.
If $\Delta t_{k}$ is negative, the impulse is blocked due to the ongoing refractory period $R^{P}\!\left(\Delta t_{k-1},\cdot,\cdot\right)$.
After an impulse is conducted, $R^{P}\!\left(\Delta t_{k},\cdot,\cdot\right)$ and $D^{P}\!\left(\Delta t_{k},\cdot,\cdot\right)$ of the current node are updated according to Eqs.~(\ref{eq:Rmod})(\ref{eq:Dmod})(\ref{eq:deltati}).
Details about how the impulses are processed chronologically and node by node, using a priority queue of nodes and sorted by impulse arrival time can be found in (\cite{Wallman2018}).
 
The input to the AV node mode is a series of atrial impulses during AF, with inter-arrival times modeled according to a Pearson Type IV distribution (\cite{Climent2011a}).
The AA series is generated with a point process with independent inter-arrival times and is completely characterized by the four parameters of the Pearson Type IV distribution, namely the mean $\mu$, standard deviation $\sigma$, skewness $\gamma$ and kurtosis $\kappa$.

The output of the AV node model is a series of ventricular impulses, where $t^{V}_{q}$ denotes the time of the $q$:th ventricular impulse.
As the refractory period $R^{P}\!\left(\Delta t_{k},\cdot,\cdot\right)$ and conduction delay $D^{P}\!\left(\Delta t_{k},\cdot,\cdot\right)$ are history-dependent, the first 1\,000 ventricular impulses leaving the AV node model are excluded from analysis to avoid transient effects.

\subsubsection{Simulation of AV nodal conduction}
\label{sec:modelAVcond}
For the training and validation, a dataset with 2 million unique parameter sets was generated.
This dataset was divided into 20 datasets with 100\,000 parameter sets each, where a dataset was either used for training or validation of one of ten realizations of the convolutional neural network (CNN) that is described in Section \ref{sec:CNNarch}.
Simulations were performed with each parameter set using the AV node model described in Section \ref{sec:networkmodel}.
For each simulation, a series of 11\,000 AA intervals was generated using the Pearson Type IV distribution, defined by the four parameters $\mu$, $\sigma$, $\gamma$, and $\kappa$.
The parameter $\mu$ was randomly drawn from ${\displaystyle {\mathcal {U}}{[100,250]}}$\,ms and $\sigma$ was randomly drawn from ${\displaystyle {\mathcal {U}}{[15,30]}}$\,ms.
The parameters $\gamma$ and $\kappa$ were kept fixed at 1 and 6, respectively, since they cannot be estimated from the f-waves of the ECG (\cite{Plappert2022}).
Negative AA intervals were excluded from the impulse series.
The model parameters for the refractory period $\boldsymbol{\theta}^{P}_{R}$ and conduction delay $\boldsymbol{\theta}^{P}_{D}$ of the SP and FP were drawn from bounded uniform distributions and the model parameters of the CN were kept fixed according to Table \ref{tab:2}.
\begin{table}
\caption{\label{tab:2} AV Node model parameters used for simulated data.}
\vspace{4 mm}
\centerline{
\begin{tabular}{lllll} \hline\hline
\multicolumn{2}{l}{Parameters} & P$\equiv$SP (ms) & P$\equiv$FP (ms) & P$\equiv$CN (ms) \\ \hline
\multirow{3}{*}{$\boldsymbol{\theta}^{P}_{R}$} & $R_{min}^{P}$ & $\displaystyle {\mathcal {U}}{[250,600]}$ & $\displaystyle {\mathcal {U}}{[250,600]}$ & $250$ \\
 & $\Delta R^{P}$ & $\displaystyle {\mathcal {U}}{[0,600]}$ & $\displaystyle {\mathcal {U}}{[0,600]}$ & $0$ \\
 & $\tau_{R}^{P}$ & $\displaystyle {\mathcal {U}}{[50,300]}$ & $\displaystyle {\mathcal {U}}{[50,300]}$ & $1$ \\
\multirow{3}{*}{$\boldsymbol{\theta}^{P}_{D}$} & $D_{min}^{P}$ & $\displaystyle {\mathcal {U}}{[0,30]}$ & $\displaystyle {\mathcal {U}}{[0,30]}$ & $0$ \\ 
 & $\Delta D^{P}$ & $\displaystyle {\mathcal {U}}{[0,75]}$ & $\displaystyle {\mathcal {U}}{[0,75]}$ & $0$ \\
 & $\tau_{D}^{P}$ & $\displaystyle {\mathcal {U}}{[50,300]}$ & $\displaystyle {\mathcal {U}}{[50,300]}$ & $1$\\ \hline\hline
\end{tabular}}
\end{table}
The given ranges were in line with our previous work (\cite{Plappert2022}).
The model parameters for the respiratory modulation and simulated respiration signal that are used in Sec. \ref{sec:respirationsignal} were also drawn from bounded uniform distributions, with $a_{resp}$ randomly drawn from ${\displaystyle {\mathcal {U}}{[-0.1,0.5]}}$, $f_{resp}$ randomly drawn from ${\displaystyle {\mathcal {U}}{[0.1,0.4]}}$\,Hz and $\eta$ randomly drawn from ${\displaystyle {\mathcal {U}}{[0.2,4]}}$.
For testing, another dataset with 2 million unique parameter sets was generated using the same ranges listed above, except for $a_{resp}$, which was randomly drawn from ${\displaystyle {\mathcal {U}}{[0,0.4]}}$.

When sampling, initially a value for $a_{resp}$ was drawn from a uniform distribution.
% The sampling was performed in the following way
%Initially a value for $a_{resp}$ was drawn from a uniform distribution.
To exclude non-physiological parameter sets from the dataset, we resampled the rest of the parameters until the following five selection criteria were met:
%We applied the following five selection criteria to exclude non-physiological parameter sets from the dataset:
1) the slow pathway in every parameter set must have a higher conduction delay 
\mbox{$D^{SP}\!\left(\Delta t_{k},\cdot,\cdot\right)>D^{FP}\!\left(\Delta t_{k},\cdot,\cdot\right)$} and lower refractory period
\mbox{$R^{SP}\!\left(\Delta t_{k},\cdot,\cdot\right)<R^{FP}\!\left(\Delta t_{k},\cdot,\cdot\right)$} than the fast pathway for all $\Delta t_{k}$;
2) the resulting average RR interval has to fall within the range of 300 ms to 1000 ms, which corresponds to heart rates between 60 bpm and 200 bpm;
3) the resulting root mean square of successive RR interval differences (RR rmssd) has to be above 100\,ms;
4) the resulting sample entropy of the RR series has to be above 1;
5) the relative contribution of the respiration frequency in the frequency spectrum of the RR series with zero-mean $F_{RR}(f_{resp})/\sum_fF_{RR}(f)$ has to be below 7\% to exclude RR series with visible periodicity.
Note that the frequency spectrum is computed from the RR series with 240 samples and the sampling rate of 4\,Hz.

Similar to the clinical data described in Section \ref{sec:ecgprocessing}, RR series were computed from intervals between the simulated ventricular impulses, and the time of each RR interval sample was set to the time of the first ventricular impulse.
The resulting non-uniformly sampled RR series were interpolated to a uniform sampling rate of 4\,Hz using piecewise cubic hermite interpolating polynomials as implemented in MATLAB ('pchip', version R2023a, RRID:SCR\_001622).
The simulated RR series were cut into \mbox{1-minute} segments of length $N=240$, resulting in RR series $\mathcal{X}_{RR}^{Sim}$ of dimension $1\!\times\!N$.
For each RR series, $\mu$ was repeated $N$ times to form a vector $\mathcal{X}_{AFR}^{Sim}$ of dimension $1\!\times\! N$, corresponding to the mean atrial arrival rate.

\subsubsection{Modelling respiratory signals}
\label{sec:respirationsignal}
For the modeling of the respiratory signals resembling joint-lead EDR signals (cf. Sec. \ref{sec:extractionjointleadEDR}), we start with the underlying assumption that respiration can be described according to $m(t)=\textrm{sin}(2\pi t f_{resp})$, i.e., by a sine wave oscillating at the respiratory frequency $f_{resp}$.
Eight identical lead-specific EDR signals $m'_{p}(t)$ with $p=1,...,8$ were created, composed of non-uniform samples of $m(t)$ at the times of the ventricular impulses $t_{q}^{V}$ generated by the AV node model.
To emulate lead-specific EDR signals, Gaussian noise with standard deviation $\eta$ was added to all samples of $m'_{p}(t)$, making them non-identical.

Next, $m'_{p}(t)$ were processed in five steps to mimic the processing steps for the clinical data (cf. Sec. \ref{sec:extractionleadspec} and \ref{sec:extractionjointleadEDR}):
1) using the same criteria as for the outlier exclusion in the clinical data, all samples in $m'_{p}(t)$ for the same ventricular impulse were excluded as outliers, if the value in one of the eight leads was outside the mean$\pm$3\,std, computed for each lead within a \mbox{1-minute} running window;
2) as for the clinical lead-specific EDR signals, the simulated signals $m'_{p}(t)$ were interpolated to a uniform sampling rate of 4\,Hz using the modified Akima algorithm as implemented in MATLAB ('makima', version R2023a, RRID:SCR\_001622), resulting in $m'_{p}(n)$;
3) $m'_{p}(n)$ were cut into \mbox{1-minute} segments of length $N=240$ and had the dimension $8\!\times\!N$;
4) the resampled and cut signals are filtered with a Butterworth highpass filter of order 4 with the cut-off frequency 0.08\,Hz to remove baseline-wander; 
5) a joint-lead EDR signal $\mathcal{X}_{Resp}^{Sim}$ with dimension $1\!\times\!N$ was extracted from $m'_{p}(n)$ using the periodic component analysis described in Section \ref{sec:extractionjointleadEDR}.

\subsection{Prediction of respiratory modulation}
\label{sec:recurrentneuralnetwork}
\subsubsection{Architecture of 1-dimensional convolutional neural network}
\label{sec:CNNarch}
\begin{figure}
    \centering
    \includegraphics[width=\textwidth]{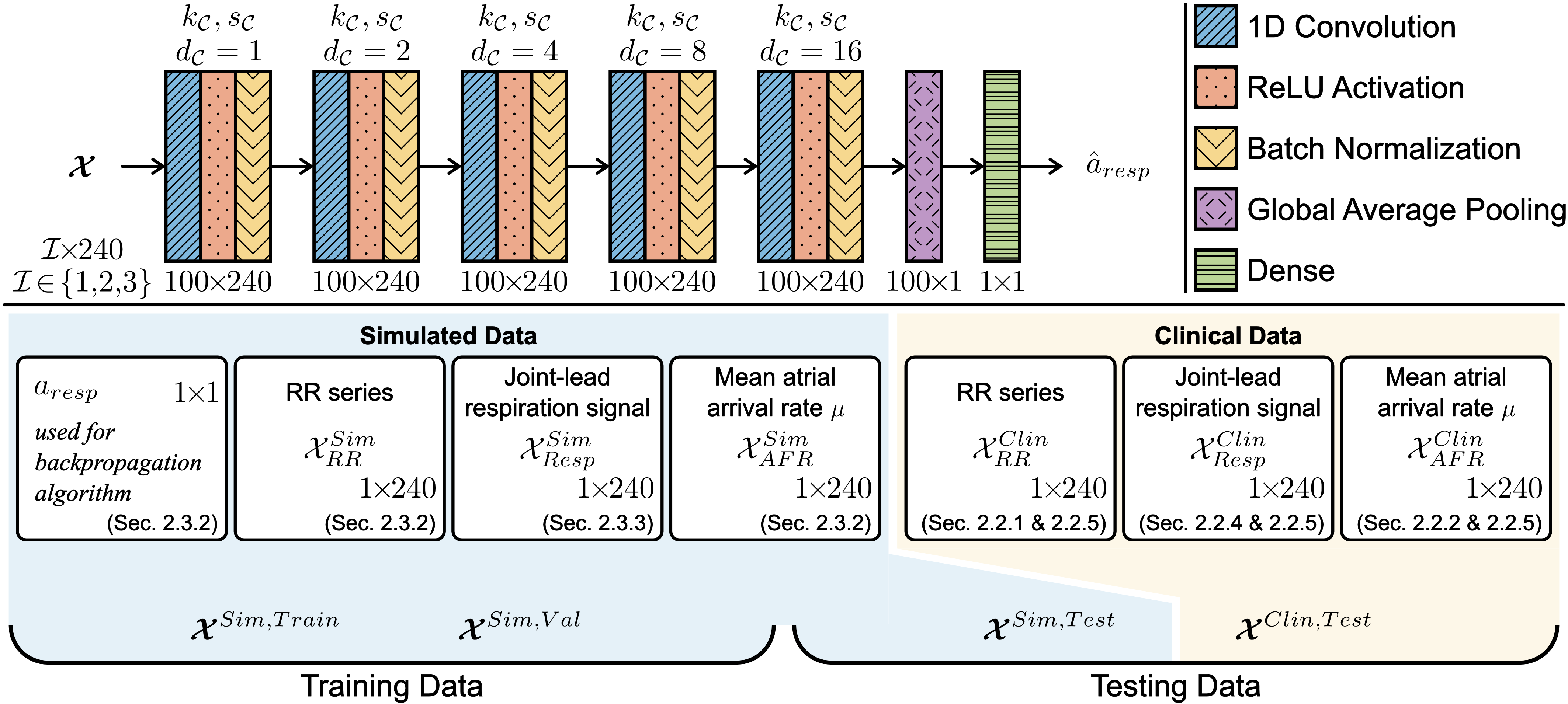}
    \caption{The CNN was composed of five 1D convolution layers with 100 filters each. The convolution layers had a kernel size $k_\mathcal{C}$, stride $s_\mathcal{C}$ and dilation factor $d_\mathcal{C}$. Training datasets $\boldsymbol{\mathcal{X}}^{Sim,Train}$, validation datasets  $\boldsymbol{\mathcal{X}}^{Sim,Val}$, and testing datasets $\boldsymbol{\mathcal{X}}^{Sim,Test}$ were constructed from the simulated data $\mathcal{X}^{Sim}_{RR}$, $\mathcal{X}^{Sim}_{Resp}$ and $\mathcal{X}^{Sim}_{AFR}$. A testing dataset $\boldsymbol{\mathcal{X}}^{Clin,Test}$ was constructed from the clinical data $\mathcal{X}^{Clin}_{RR}$, $\mathcal{X}^{Clin}_{Resp}$ and $\mathcal{X}^{Clin}_{AFR}$.}
    \label{fig:2}
\end{figure}

To predict the peak-to-peak amplitude of the respiratory modulation, $a_{resp}$, a 1D-CNN architecture was used as illustrated in Figure \ref{fig:2}.
The CNN architecture consists of five convolution layers, where each layer $l$ was composed of 100 1D-CNN filters with kernel size $k_\mathcal{C}=3$, stride $s_\mathcal{C}=1$ and dilation factor $d_\mathcal{C}=2^{l-1}$, followed by a rectified linear unit (RELU) and a batch normalization layer.
After the five convolution layers, the data passed through a global average pooling layer and dense layer, the output of which is a prediction $\hat{a}_{resp}$.
To assess the performance of the CNN with or without the RR series, respiration signal, and mean $\mu$ of the AA series, seven versions of the CNN were trained.
The respective CNNs and their input data are given as follows: the CNN $\mathcal{C}_{RR}$ was trained on the input data with the format $\boldsymbol{\mathcal{X}}=\mathcal{X}_{RR}^{Sim}$; $\mathcal{C}_{Resp}$ was trained on $\boldsymbol{\mathcal{X}}=\mathcal{X}_{Resp}^{Sim}$; $\mathcal{C}_{AFR}$ was trained on $\boldsymbol{\mathcal{X}}=\mathcal{X}_{AFR}^{Sim}$; $\mathcal{C}_{RR,Resp}$ was trained on $\boldsymbol{\mathcal{X}}=[\mathcal{X}_{RR}^{Sim}; \mathcal{X}_{Resp}^{Sim}]$; $\mathcal{C}_{RR,AFR}$ was trained on $\boldsymbol{\mathcal{X}}=[\mathcal{X}_{RR}^{Sim}; \mathcal{X}_{AFR}^{Sim}]$; $\mathcal{C}_{Resp,AFR}$ was trained on $\boldsymbol{\mathcal{X}}=[\mathcal{X}_{Resp}^{Sim}; \mathcal{X}_{AFR}^{Sim}]$; and $\mathcal{C}_{RR,Resp,AFR}$ was trained on $\boldsymbol{\mathcal{X}}=[\mathcal{X}_{RR}^{Sim}; \mathcal{X}_{Resp}^{Sim}; \mathcal{X}_{AFR}^{Sim}]$.

\subsubsection{Training the convolutional neural network}
\label{sec:trainedRNN}
For each CNN version, i.e., $\mathcal{C}_{RR}$, $\mathcal{C}_{Resp}$, $\mathcal{C}_{AFR}$, $\mathcal{C}_{RR,Resp}$, $\mathcal{C}_{RR,AFR}$, $\mathcal{C}_{Resp,AFR}$ and $\mathcal{C}_{RR,Resp,AFR}$, described in Sec. \ref{sec:CNNarch}, ten realizations were trained with unique training and validation datasets, $\boldsymbol{\mathcal{X}}^{Sim,Train}$ and $\boldsymbol{\mathcal{X}}^{Sim,Val}$, respectively, containing 100\,000 parameter sets each, as described in Sec. \ref{sec:modelAVcond}.
The CNNs were trained to estimate the $a_{resp}$ and the weights of the CNN were updated during backpropagation based on the root-mean-square error (RMSE) of the residuals.
Every epoch, $\boldsymbol{\mathcal{X}}^{Sim,Train}$ was randomly divided into 20 mini-batches, each containing input data for 5\,000 different parameter sets.
A cyclical learning rate was set for the training, where the learning rate started at $5\!\cdot\!10^{-3}$ and was increased and decreased in a 'zig-zag' between $[2\!\cdot\!10^{-3}$, $3\!\cdot\!10^{-3}$, $5\!\cdot\!10^{-3}$, $8\!\cdot\!10^{-3}$, $10\!\cdot\!10^{-3}]$ every time the RMSE of $\boldsymbol{\mathcal{X}}^{Sim,Val}$ did not improve for 50 epochs (\cite{Smith2017}).
The initial learning rate and the minimum and maximum boundary values of the cyclical learning rates were determined using the 'learning rate range test', described in (\cite{Smith2017}).
The network was validated after every epoch.
The CNN was trained until the RMSE of $\boldsymbol{\mathcal{X}}^{Sim,Val}$ did not improve for 50 epochs for each of the five learning rates and the network weights giving the lowest validation RMSE was chosen.
The prediction $\hat{a}_{resp}$ was computed as the average of the individual predictions of each of the ten CNN realizations.

\subsubsection{Prediction of respiratory modulation in simulated data}
\label{sec:predictionmodulationsimdata}
The performance of the CNN on simulated data was assessed for $\mathcal{C}_{RR}$, $\mathcal{C}_{Resp}$, $\mathcal{C}_{AFR}$, $\mathcal{C}_{RR,Resp}$, $\mathcal{C}_{RR,AFR}$, $\mathcal{C}_{Resp,AFR}$ and $\mathcal{C}_{RR,Resp,AFR}$, using the testing dataset $\boldsymbol{\mathcal{X}}^{Sim,Test}$ described in Sec. \ref{sec:modelAVcond}.
The total performance on $\boldsymbol{\mathcal{X}}^{Sim,Test}$ was assessed using the RMSE, Pearson correlation, and coefficient of determination $R^{2}$ between the true $a_{resp}$ and predicted $\hat{a}_{resp}$. 

In addition, the performance was assessed over a range of respiration frequencies $f_{resp}$ and characteristics of non-periodicity in the respiration signal $\epsilon(\mathbf{w},\tau,\mathbf{Z})$, here denoted $\epsilon$.
To produce local RMSE estimates $\sigma(f'_{resp},\epsilon')$ for specific values $f_{resp}'$ and $\epsilon'$, the following three steps were applied: 1) a squared difference $\left(a_{resp}-\hat{a}_{resp}\right)^2$ was computed for each of the 2 million parameter sets in $\boldsymbol{\mathcal{X}}^{Sim,Test}$; 2) a weighted average of the 2 million squared differences was computed using a Gaussian kernel centered at $f_{resp}'$ and $\epsilon'$ with the standard deviation of $0.015$\,Hz and $0.075$ for the $f_{resp}$ and $\epsilon$, respectively; 3) the square root of the weighted average resulted in $\sigma(f'_{resp},\epsilon')$.

In the present study, all versions of the CNN were trained and tested using 1-minute segments, with one exception: An additional CNN $\mathcal{C}_{RR,Resp,AFR}^{2.5min}$ was trained and tested using \mbox{$\boldsymbol{\mathcal{X}}^{\ast}=[\mathcal{X}_{RR}^{Sim,2.5}; \mathcal{X}_{Resp}^{Sim,2.5}; \mathcal{X}_{AFR}^{Sim,2.5}]$} containing 2.5-minute-long segments to investigate the impact of segment length on the RMSE.
For $\mathcal{C}_{RR,Resp,AFR}^{2.5min}$, ten realizations were trained with additional unique training and validation datasets, $\boldsymbol{\mathcal{X}}^{\ast Sim,Train}$ and $\boldsymbol{\mathcal{X}}^{\ast Sim,Val}$, respectively, containing 100\,000 parameter sets each.
Apart from the different segment lengths, the additional datasets were generated as described in Sec. \ref{sec:modelAVcond}.

\subsubsection{Prediction of respiratory modulation in clinical data}
\label{sec:predictionmodulationclindata}
The CNN $\mathcal{C}_{RR,Resp,AFR}$ was used for predicting $a_{resp}$ in the clinical deep breathing test data, described in \ref{sec:clinicaldata}.
The clinical predictions were used to investigate differences in $\hat{a}_{resp}$ between deep breathing and normal breathing using Monte Carlo sampling.
Using these samples, the probabilities of the following three scenarios were computed for each patient: 1) the highest $\hat{a}_{resp}$ was achieved for deep breathing, 2) the lowest $\hat{a}_{resp}$ was achieved for deep breathing and 3) the highest and lowest $\hat{a}_{resp}$ did not correspond to deep breathing.
To draw the samples for each \mbox{1-minute} segment in $\boldsymbol{\mathcal{X}}^{Clin,Test}$, the prediction $\hat{a}_{resp}$ was determined using the CNN $\mathcal{C}_{RR,Resp,AFR}$, while the $f'_{resp}$ and $\epsilon'$ were estimated by the $\hat{f}_{resp}$ and $\epsilon(\mathbf{w},\tau,\mathbf{Z})$ described in Sec. \ref{sec:extractionjointleadEDR}.
Next, values of $\hat{a}_{resp}$ were resampled 100\,000 times for each \mbox{1-minute} segment in $\mathcal{S}_{seg}^{\ast}$.
The samples were drawn from Gaussian distributions with $\hat{a}_{resp}$ as mean and $\sigma(f'_{resp},\epsilon')$ described in Sec. \ref{sec:predictionmodulationsimdata} as standard deviation.

\section{Results}
\subsection{Analysis of clinical data}
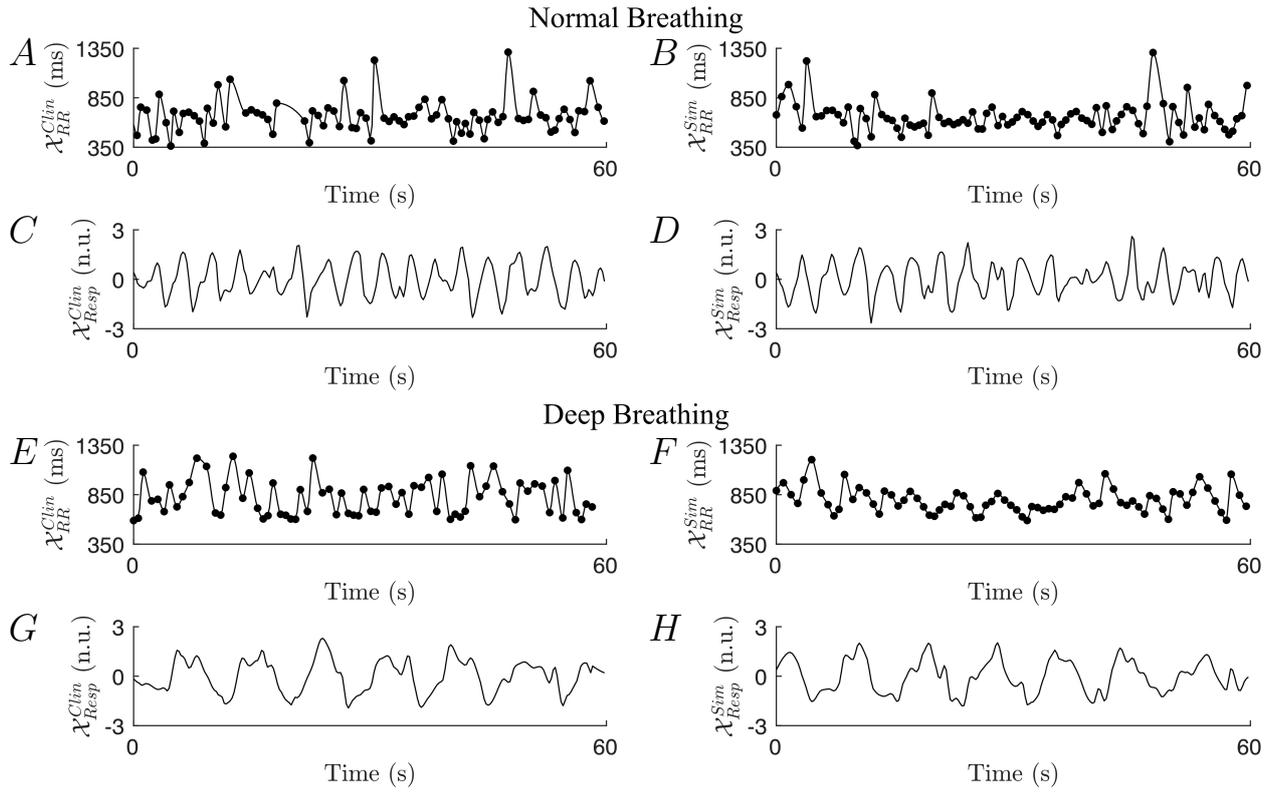
\begin{figure}
    \centering
    \begin{overpic}[width=\textwidth]{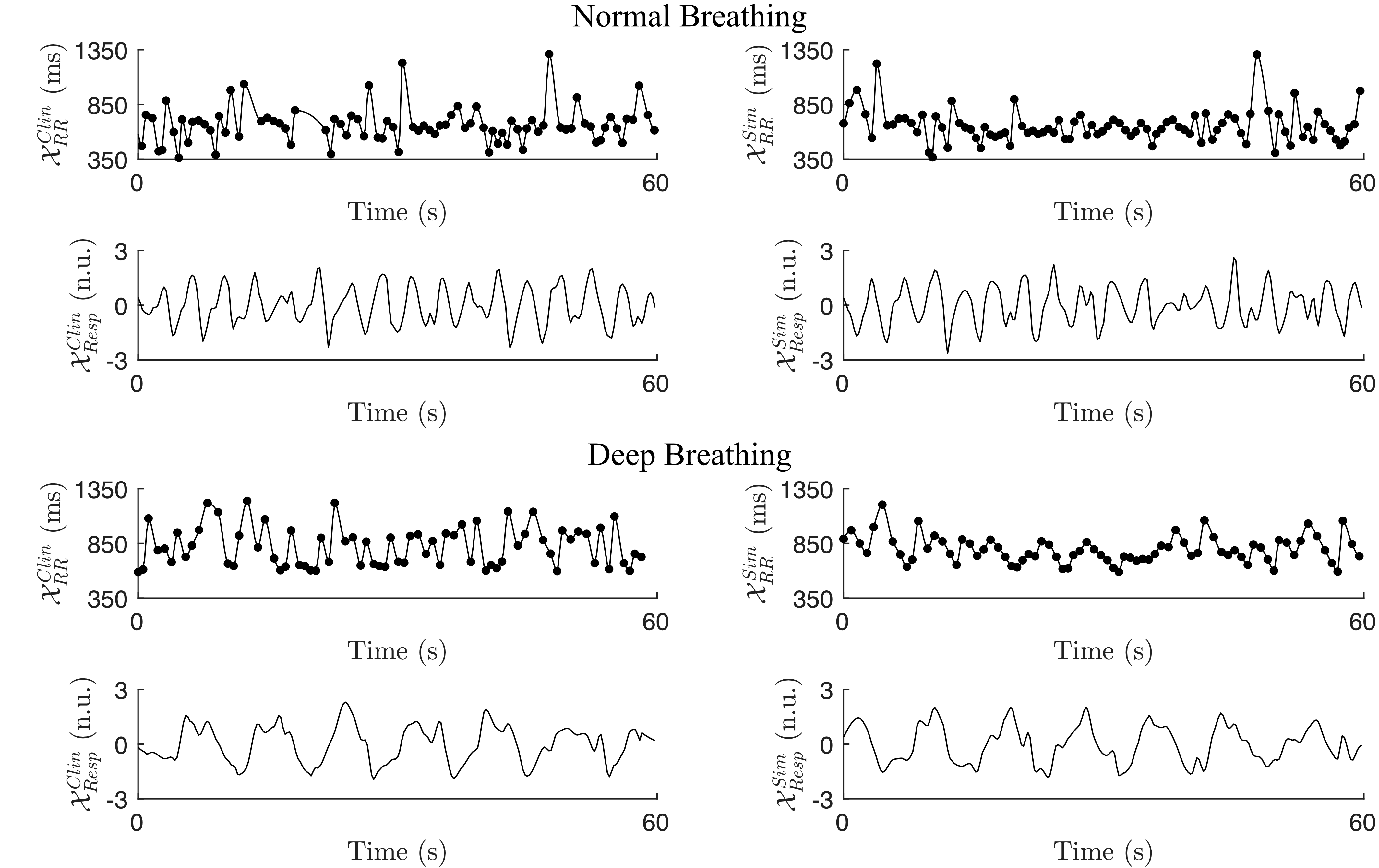}
    \put (0,58) {\Large$A$}
    \put (51,58) {\Large$B$}
    \put (0,43.8) {\Large$C$}
    \put (51,43.8) {\Large$D$}
    \put (0,26) {\Large$E$}
    \put (51,26) {\Large$F$}
    \put (0,11.8) {\Large$G$}
    \put (51,11.8) {\Large$H$}
    \end{overpic}
    \caption{Two examples of clinical RR series (A+E), simulated RR series (B+F), clinical respiration signals (C+G), and simulated respiration signals (D+H) during normal breathing (A-D) and deep breathing (E-H).}
    \label{fig:3}
\end{figure}

\begin{figure}
    \centering
    \includegraphics{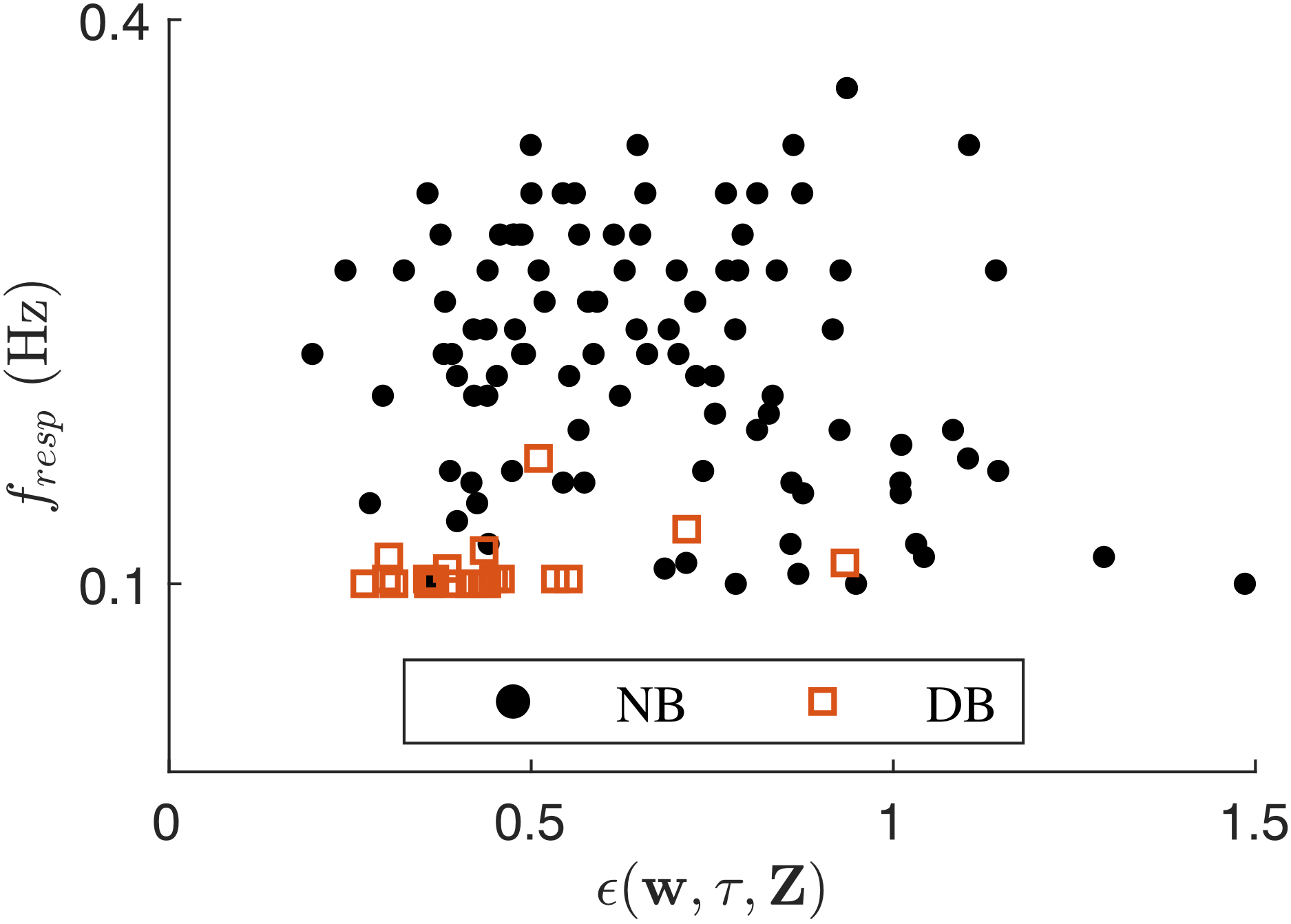}
    \caption{Scatter plot showing $\epsilon(\mathbf{w},\tau,\mathbf{Z})$ over $\hat{f}_{resp}$ for each \mbox{1-minute} segment during normal breathing (NB) and deep breathing (DB).}
    \label{fig:4}
\end{figure}

The length of the interpolated RR series varied between patients depending on the duration of the recordings; during normal breathing, the length of the RR series was in the range between 288\,s and 328\,s; during deep breathing, the length of the RR series was in the range between 57\,s and 72\,s.
Statistics quantifying the clinical dataset are shown in Table \ref{tab:3}.
\begin{table}
\caption{\label{tab:3} Characteristics of clinical and simulated data. The training data is divided into 20 datasets with equal size to train the 10 realizations of the CNN with unique $\boldsymbol{\mathcal{X}}^{Sim,Train}$ and $\boldsymbol{\mathcal{X}}^{Sim,Val}$. The variables $\overline{AFR}$, $f_{resp}$, and $a_{resp}$ characterize estimates in the clinical data and model parameters in the simulated data.}
\vspace{4 mm}
\centerline{\begin{tabular}{lcccc} \hline\hline
 & \multicolumn{2}{c}{Clinical data $\boldsymbol{\mathcal{X}}^{Clin,Test}$} & \multicolumn{2}{c}{Simulated data}  \\
 & Normal breathing & Deep breathing & Training Data & Testing Data \\ 
 & & & $[\boldsymbol{\mathcal{X}}^{Sim,Train};\boldsymbol{\mathcal{X}}^{Sim,Val}]$ & $\boldsymbol{\mathcal{X}}^{Sim,Test}$ \\ \hline
Number of $\boldsymbol{\mathcal{X}}$ &$98$&$22$&$10\cdot2\cdot100\,000$&$2\,000\,000$\\
RR mean (ms) &$763\pm173$&$747\pm162$&$676\pm164{}^{\dagger}$&$676\pm164{}^{\dagger}$\\
RR rmssd (ms) &$262\pm100$&$230\pm60$&$188\pm60{}^{\dagger,\ddagger}$&$185\pm58{}^{\dagger,\ddagger}$\\
RR sample entropy &$2.08\pm0.49$&$2.18\pm0.63$&$1.53\pm0.39{}^{\dagger,\ddagger}$&$1.52\pm0.38{}^{\dagger,\ddagger}$\\
$F_{RR}(f_{resp})/\sum_fF_{RR}(f)$ (\%) &$2.5\pm1.3\,$&$1.1\pm0.8\,$&$3.4\pm1.8\,{}^{\dagger,\ddagger}$&$3.3\pm1.7\,{}^{\dagger,\ddagger}$\\
$\overline{AFR}$ (Hz) &$6.99\pm0.7$&$6.95\pm0.71$&$5.97\pm1.57{}^{\dagger,\ddagger}$&$5.96\pm1.57{}^{\dagger,\ddagger}$\\
$f_{resp}$ (Hz) &$0.220\pm0.067$&$0.107\pm0.015$&$0.263\pm0.085{}^{\dagger,\ddagger}$&$0.261\pm0.085{}^{\dagger,\ddagger}$\\
$\epsilon$ &$0.66\pm0.25$&$0.44\pm0.15$&$0.64\pm0.27{}^{\ddagger}$&$0.64\pm0.27{}^{\ddagger}$\\
$a_{resp}$ &$0.282\pm0.101$&$0.285\pm0.131$&$0.200\pm0.173{}^{\dagger,\ddagger}$&$0.200\pm0.115{}^{\dagger,\ddagger}$\\
\hline\hline
\multicolumn{5}{l}{${}^{\dagger}p<0.05$ vs normal breathing. ${}^{\ddagger}p<0.05$ vs deep breathing.}
\end{tabular}}
\end{table}
In accordance with the exclusion criteria defined in Sec. \ref{sec:PostprocessingRRseriesEDRsignals}, 98 out of 120 non-overlapping \mbox{1-minute} segments remained in the normal breathing data and 22 out of 28 \mbox{1-minute} segments remained in the deep breathing data.
Typical examples of a clinical RR series $\mathcal{X}_{RR}^{Clin}$ and joint-lead respiration signal $\mathcal{X}_{Resp}^{Clin}$ during normal breathing and deep breathing, respectively, are shown in Fig. \ref{fig:3}.
The characteristics of these signals, listed in Table \ref{tab:4} are within 1 standard deviation of the population mean (cf. Table \ref{tab:3}).
\begin{table}
\caption{\label{tab:4} Characteristics of the clinical and simulated examples are shown in Fig. \ref{fig:3}.}
\vspace{4 mm}
\centerline{\begin{tabular}{lccccccc} \hline\hline
Signals & RR mean (ms) & RR rmssd (ms) & RR sample entropy & $a_{resp}$ & $f_{resp}$ (Hz) & $\eta$ & $\epsilon(\mathbf{w},\tau,\mathbf{Z})$ \\ \hline
A/C & 661 & 250 & 1.85 & - & 0.286 & - & 0.47 \\
B/D & 651 & 204 & 1.91 & 0.36 & 0.288 & 2.48 & 0.76 \\
E/G & 818 & 251 & 2.28 & - & 0.118 & - & 0.44 \\
F/H & 792 & 138 & 1.97 & 0.05 & 0.116 & 1.46 & 0.45 \\
\hline\hline\\
\end{tabular}}
\end{table}
Fluctuations in the clinical RR series matching the respiration frequencies were not clearly visible and $F_{RR}(f_{resp})/\sum_fF_{RR}(f)$ was always below 7\%.
The respiration signals estimated from clinical data had $\epsilon(\mathbf{w},\tau,\mathbf{Z})$ ranging between 0.198 and 1.485.
The clinical value pairs of $\epsilon(\mathbf{w},\tau,\mathbf{Z})$ and respiration frequency $\hat{f}_{resp}$ are shown in Fig. \ref{fig:4}.
There was a statistically significant weak negative correlation between $\hat{f}_{resp}$ and $\epsilon(\mathbf{w},\tau,\mathbf{Z})$ in the clinical data during normal breathing ($r=-.217, p=0.032$), but no significant correlation during deep breathing.

\subsection{Simulated RR series and respiration signals}
\label{sec:simRRRespSignals}
The statistics quantifying $\boldsymbol{\mathcal{X}}^{Sim,Train}$, $\boldsymbol{\mathcal{X}}^{Sim,Val}$ and $\boldsymbol{\mathcal{X}}^{Sim,Test}$ are shown in Table \ref{tab:3} together with $\boldsymbol{\mathcal{X}}^{Clin,Test}$.
The simulated datasets were created according to the description in Sec. \ref{sec:simulateddata} and compared to the clinical data using the unpaired t-test.
It should be noted that although there are significant differences between the characteristics of the clinical and simulated data, the distributions of the simulated data cover the distribution of the clinical data.
The heart rate was on average slightly faster and more regular in $\boldsymbol{\mathcal{X}}^{Sim}$ than in $\boldsymbol{\mathcal{X}}^{Clin}$, as indicated by the differences in \mbox{RR mean}, \mbox{RR rmssd}, and \mbox{RR sample entropy}.
Further, the RR series in $\boldsymbol{\mathcal{X}}^{Sim}$ showed on average more visible fluctuations matching the respiration frequency compared to the RR series in $\boldsymbol{\mathcal{X}}^{Clin}$, as indicated by the difference in $F_{RR}(f_{resp})/\sum_fF_{RR}(f)$.
The $\overline{AFR}$ was on average slightly lower in $\boldsymbol{\mathcal{X}}^{Sim}$ than in $\boldsymbol{\mathcal{X}}^{Clin}$, whereas $f_{resp}$ was slightly higher.
In normal breathing, $\epsilon$ in $\boldsymbol{\mathcal{X}}^{Clin}$ was comparable to $\boldsymbol{\mathcal{X}}^{Sim}$; however, in deep breathing, $\epsilon$ was lower in $\boldsymbol{\mathcal{X}}^{Clin}$ than in $\boldsymbol{\mathcal{X}}^{Sim}$.

Examples of a simulated RR series $\mathcal{X}_{RR}^{Sim}$ and joint-lead respiration signal $\mathcal{X}_{Resp}^{Sim}$ resembling clinical signals during normal breathing and deep breathing, respectively, are shown in Fig. \ref{fig:3}.
The signals were chosen based on similarities to the clinical signals in the RR series characteristics and respiration signal morphology.
The characteristics of these signals are listed in Table \ref{tab:4}.
Note, that while the peak-to-peak amplitude of respiratory modulation $a_{resp}$ is high during normal breathing and low during deep breathing in this example, a general conclusion about the $a_{resp}$ values of the clinical signals can not be drawn from this comparison and is not intended.
When emulating lead-specific EDR signals and adding Gaussian noise with standard deviation $\eta$, the simulated data showed a strong correlation between $\eta$ and $\epsilon$ ($\rho=0.89$, $p<10^{-5}$).
The examples in Fig. \ref{fig:3} are representative of this correlation with the $\eta$ and $\epsilon$ listed in Table \ref{tab:4}, where $\mathcal{X}_{Resp}^{Sim}$ in Fig. \ref{fig:3}D was generated with a higher $\eta$ and showed a higher $\epsilon$ compared to $\mathcal{X}_{Resp}^{Sim}$ in Fig. \ref{fig:3}H.

\subsection{Accuracy of convolutional neural network}
\begin{figure}
    \centering
    \includegraphics{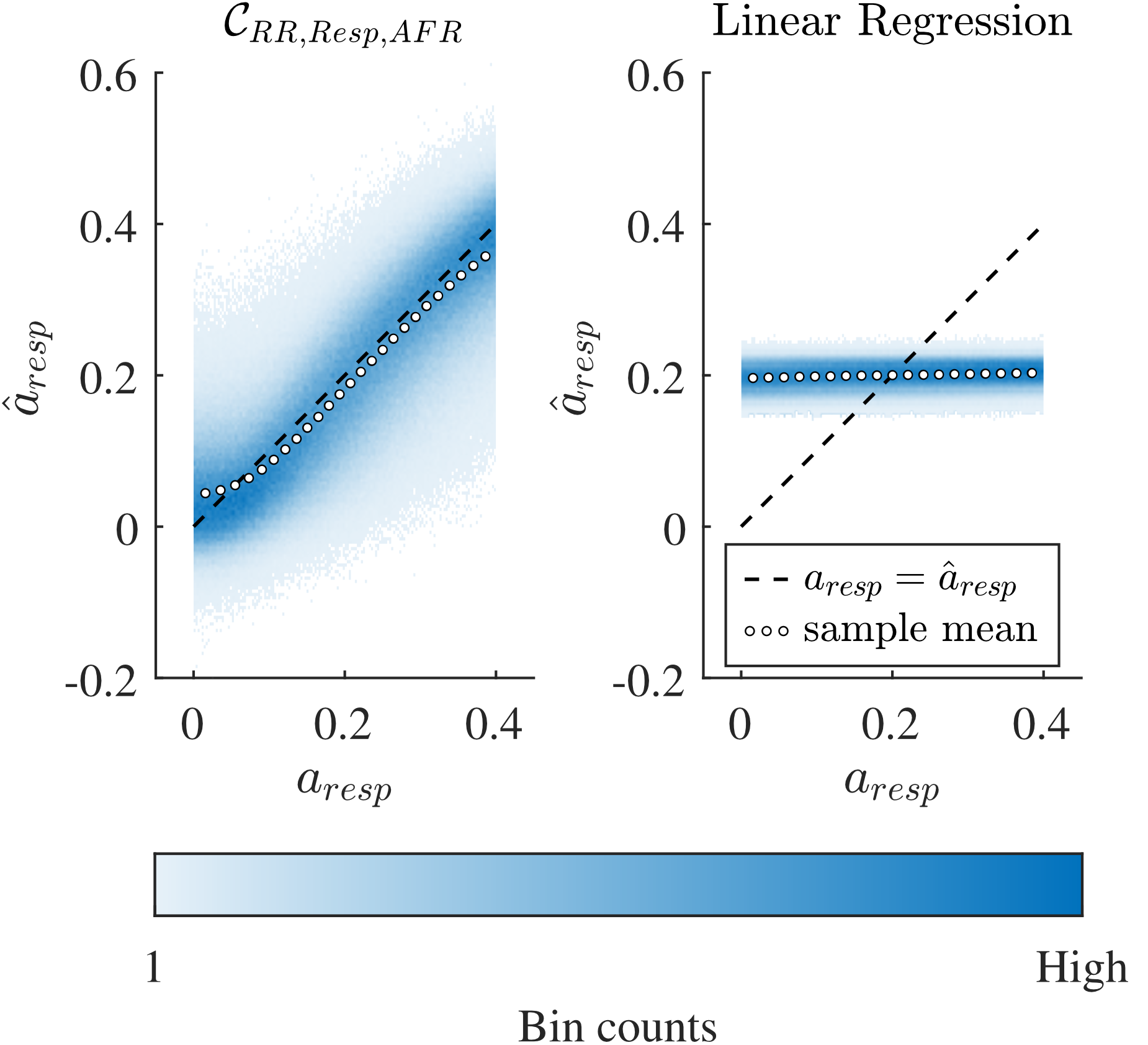}
    \caption{Binned scatter plot of predicted $\hat{a}_{resp}$ versus true $a_{resp}$ for the CNN $\mathcal{C}_{RR,Resp,AFR}$ and linear regression $\mathcal{L}_{RR,Resp,AFR}$, where both were based on the same input data \mbox{$\boldsymbol{\mathcal{X}}=[\mathcal{X}_{RR}^{Sim}; \mathcal{X}_{Resp}^{Sim}; \mathcal{X}_{AFR}^{Sim}]$}. The black dotted line shows where $\hat{a}_{resp}$ is equal to $a_{resp}$. The white dotted line shows the sample mean of the $\hat{a}_{resp}$ prediction.}
    \label{fig:5}
\end{figure}

All CNNs $\mathcal{C}_{RR}$, $\mathcal{C}_{Resp}$, $\mathcal{C}_{AFR}$, $\mathcal{C}_{RR,Resp}$, $\mathcal{C}_{RR,AFR}$, $\mathcal{C}_{Resp,AFR}$ and $\mathcal{C}_{RR,Resp,AFR}$, described in Sec. \ref{sec:CNNarch} and trained according to Sec. \ref{sec:trainedRNN}, were tested using $\boldsymbol{\mathcal{X}}^{Sim,Test}$ described in Sec. \ref{sec:modelAVcond}.
The resulting distribution of predicted $\hat{a}_{resp}$ over true $a_{resp}$ for $\mathcal{C}_{RR,Resp,AFR}$ is shown in Fig. \ref{fig:5}.
The corresponding distribution for a prediction using linear regression $\mathcal{L}_{RR,Resp,AFR}$ based on the same data $\boldsymbol{\mathcal{X}}=[\mathcal{X}_{RR}^{Sim}; \mathcal{X}_{Resp}^{Sim}; \mathcal{X}_{AFR}^{Sim}]$ is also displayed in Fig. \ref{fig:5} for comparison.
The RMSE, Pearson sample correlation and $R^{2}$ are listed for the seven CNN versions and $\mathcal{L}_{RR,Resp,AFR}$ in Table \ref{tab:5}.
\begin{table}
\caption{\label{tab:5} RMSE, Pearson sample correlation and $R^{2}$ of the seven CNN versions and linear regression $\mathcal{L}_{RR,Resp,AFR}$ using 1-minute segments, and $\mathcal{C}_{RR,Resp,AFR}^{2.5min}$ using 2.5-minute segments.}
\vspace{4 mm}
\centerline{\begin{tabular}{lccc} \hline\hline
CNN version & RMSE & Pearson correlation $r$ & $R^{2}$ \\ \hline
$\mathcal{C}_{RR,Resp,AFR}^{2.5min}$ & $0.050$ & $0.923$ & $0.816$\\
$\mathcal{C}_{RR,Resp,AFR}$ &$0.066 $& $0.855$ & $0.674$\\
$\mathcal{C}_{RR,Resp}$ &$0.070 $& $0.830$ & $0.636$\\
$\mathcal{C}_{RR,AFR}$ &$0.070 $& $0.837$ & $0.630$\\
$\mathcal{C}_{RR}$ &$0.074 $& $0.805$ & $0.585$\\
$\mathcal{C}_{Resp,AFR}$ &$0.098 $& $0.583$ & $0.284$\\
$\mathcal{C}_{Resp}$ &$0.101 $& $0.513$ & $0.231$\\
$\mathcal{C}_{AFR}$ &$0.115 $& $0.073$ & $0.001$\\
$\mathcal{L}_{RR,Resp,AFR}$ & $0.114$ & $0.131$ & $0.017$ \\
\hline\hline\\
\end{tabular}}
\end{table}
The $\mathcal{C}_{RR,Resp,AFR}$ resulted in the lowest RMSE and highest correlation and $R^{2}$.
The CNNs $\mathcal{C}_{AFR}$, $\mathcal{C}_{Resp}$ and $\mathcal{C}_{Resp,AFR}$ without RR series in the input data performed poorly.
The $\mathcal{C}_{RR}$ predicted $\hat{a}_{resp}$ with an RMSE of 0.074, where the addition of $\mathcal{X}_{Resp}^{Sim}$ or $\mathcal{X}_{AFR}^{Sim}$ to the input improved the accuracy of the $\hat{a}_{resp}$ prediction slightly.

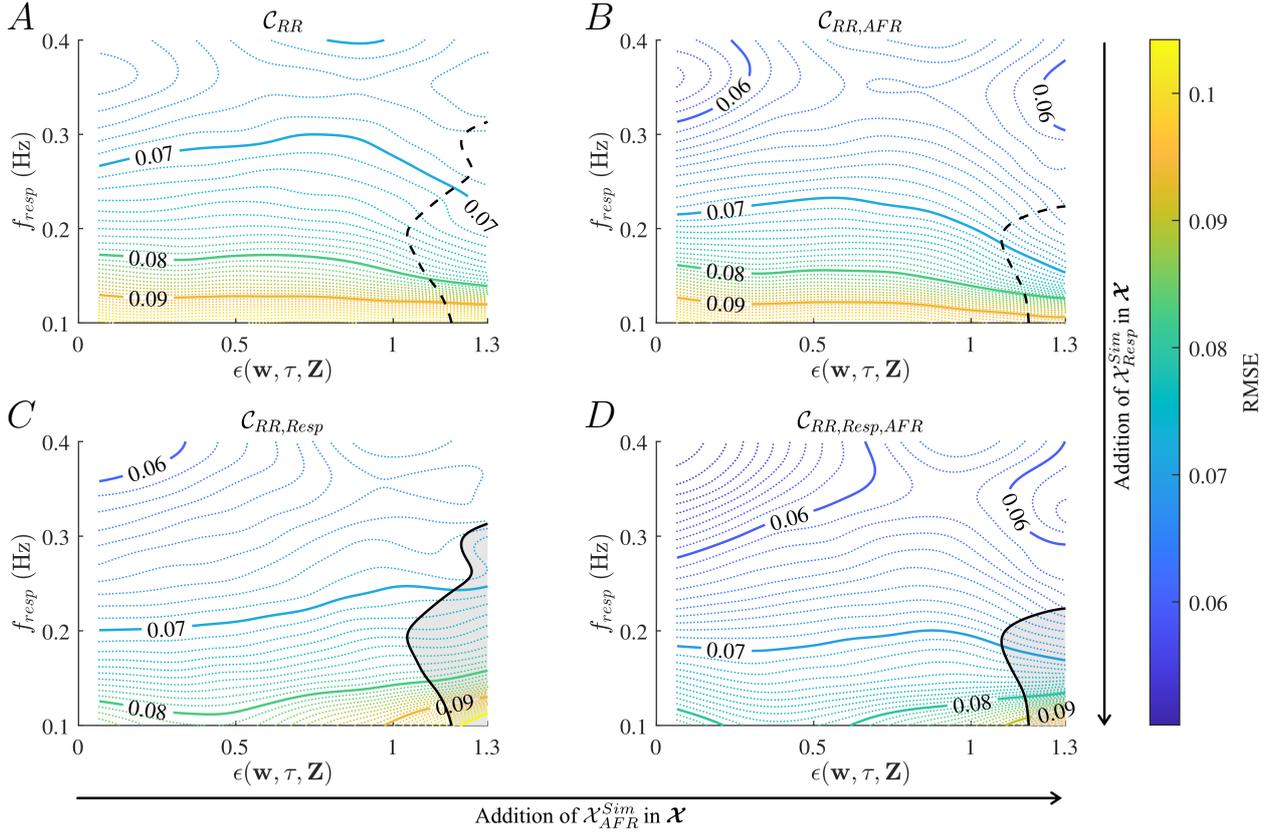
\begin{figure}
    \centering
    \begin{overpic}[width=\textwidth]{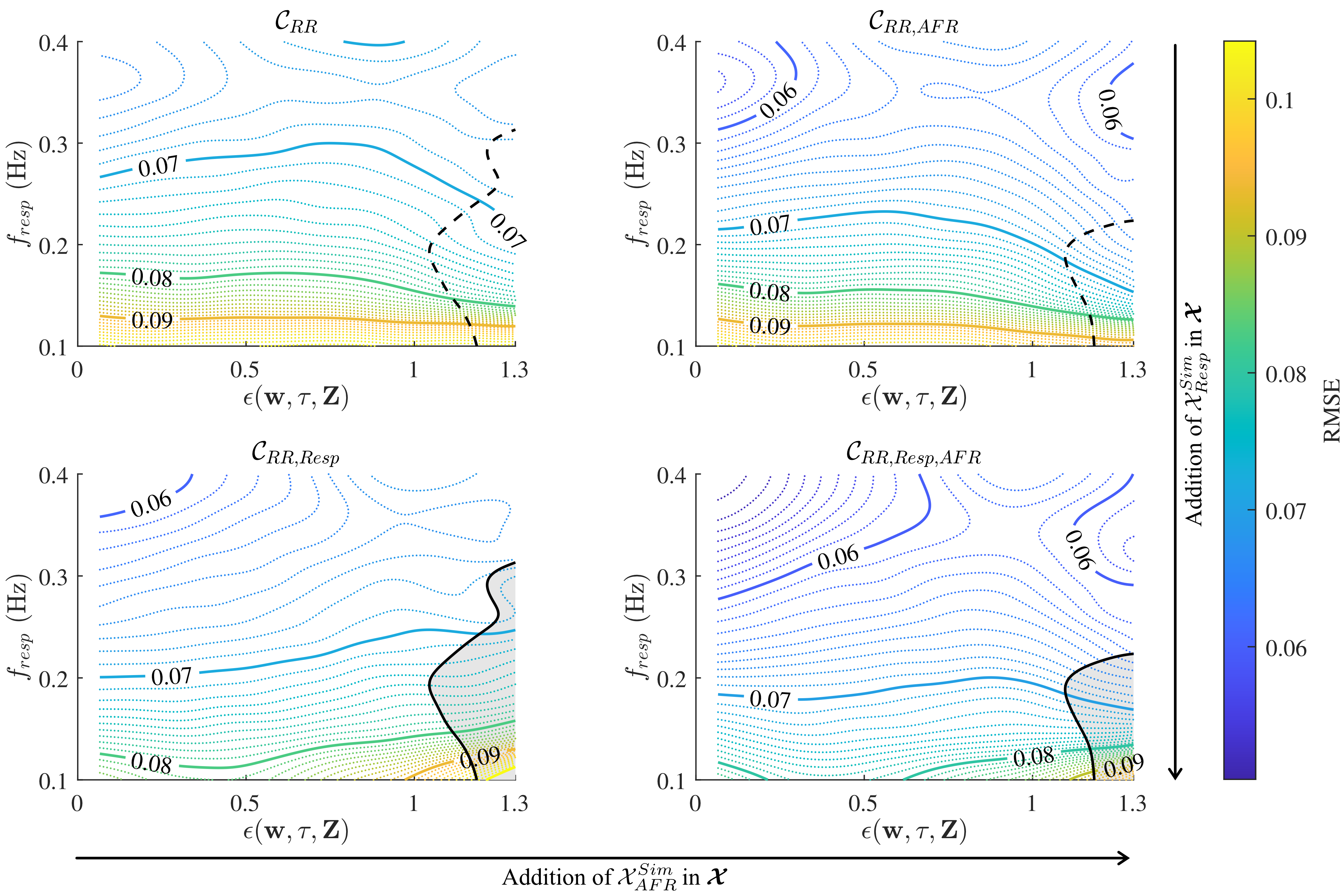}
    \put (0,64.5) {\Large$A$}
    \put (46,64.5) {\Large$B$}
    \put (0,32.5) {\Large$C$}
    \put (46,32.5) {\Large$D$}
    \end{overpic}
    \caption{Contour plot of local RMSE estimates over a range of $f'_{resp}$ and $\epsilon'$ for $\mathcal{C}_{RR}$ (A), $\mathcal{C}_{RR,AFR}$ (B), $\mathcal{C}_{RR,Resp}$ (C) and $\mathcal{C}_{RR,Resp,AFR}$ (D). Except for the grey region, the CNNs in C and D have a higher accuracy than the CNNs in A and B, respectively.}
    \label{fig:6}
\end{figure}

For $\mathcal{C}_{RR}$, $\mathcal{C}_{RR,AFR}$, $\mathcal{C}_{RR,Resp}$ and $\mathcal{C}_{RR,Resp,AFR}$, the local RMSE of $\hat{a}_{resp}$ for specific $f_{resp}'$ and $\epsilon'$ were computed according to Sec. \ref{sec:predictionmodulationsimdata} and illustrated in Fig. \ref{fig:6}.
It can be seen in all four contour plots that the RMSE is dependent on $f'_{resp}$ and $\epsilon'$. 
The CNNs produce more accurate predictions for data with a high $f'_{resp}$ and low $\epsilon'$, however, the RMSE is more sensitive to changes in $f'_{resp}$.
Adding $\mathcal{X}_{AFR}^{Sim}$ to the input improves the RMSE for all values of $f_{resp}$ and $\epsilon$.
While the addition of $\mathcal{X}_{Resp}^{Sim}$ to the input improves the RMSE for most $f_{resp}'$ and $\epsilon'$, it worsens the RMSE for high $\epsilon'$ and low $f'_{resp}$ as indicated in Fig. \ref{fig:6}.
Within the indicated region, the accuracy of $\hat{a}_{resp}$ is higher without $\mathcal{X}_{Resp}^{Sim}$ in the input data.

The accuracy of the CNN improves with longer input data, indicated by the fact that the RMSE of $\mathcal{C}_{RR,Resp,AFR}^{2.5min}$ was $0.050$.
The RMSE, Pearson sample correlation and $R^{2}$ is listed for $\mathcal{C}_{RR,Resp,AFR}^{2.5min}$ in Table \ref{tab:5}.
The RMSE improved for all values of $\epsilon'$ and $f'_{resp}$, whereas the local RMSE improved especially at lower $f'_{resp}$ (data not shown).

\subsection{Prediction of respiratory modulation in clinical data}
\begin{figure}
    \centering
    \begin{overpic}[width=\textwidth]{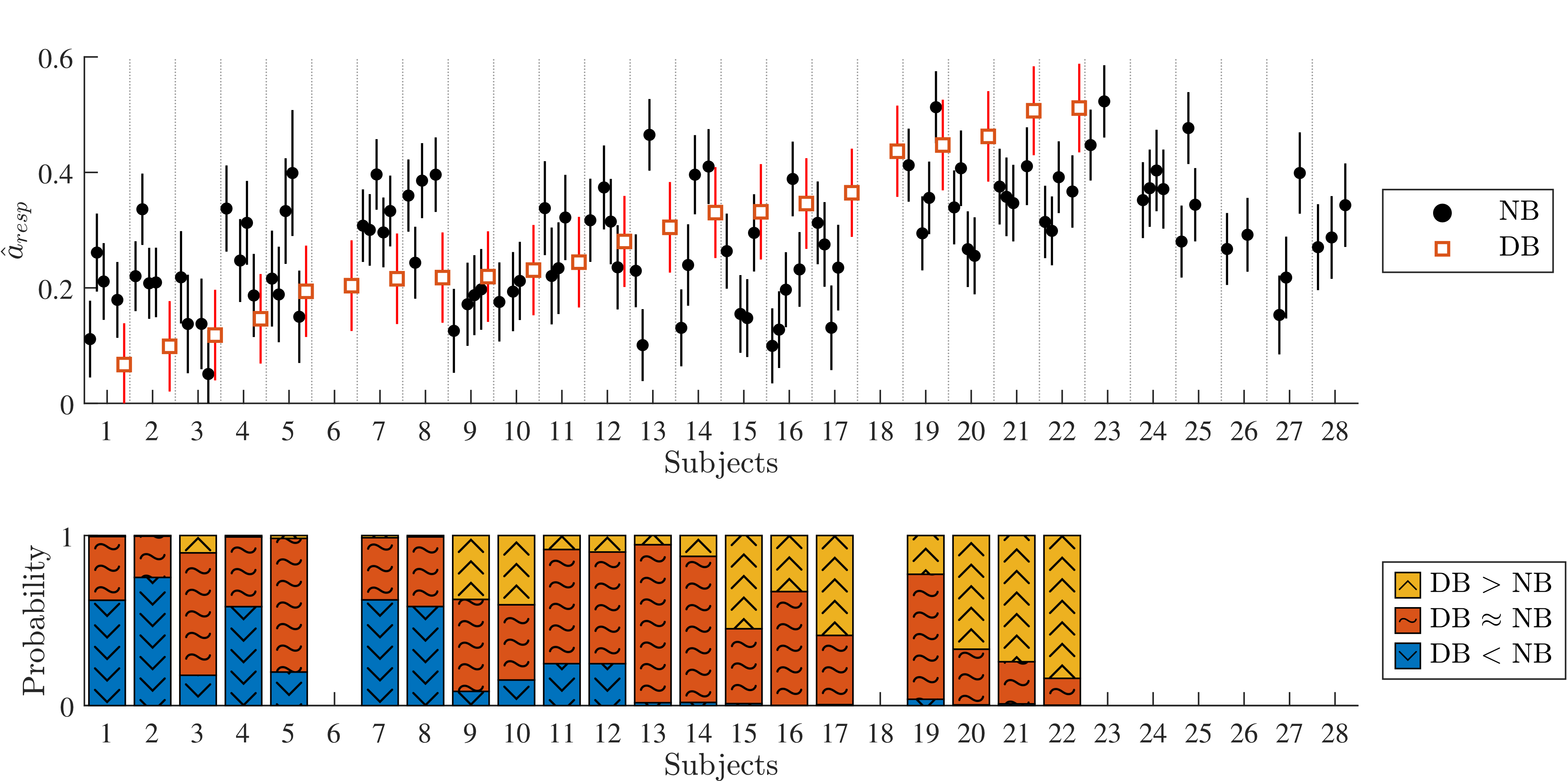}
    \put (0,47) {\Large$A$}
    \put (0,16.5) {\Large$B$}
    \end{overpic}
    \caption{\textbf{(A)} Black dots correspond to the predicted $\hat{a}_{resp}$ of \mbox{1-minute} segments during normal breathing (NB) and red squares correspond to $\hat{a}_{resp}$ of \mbox{1-minute} segments during deep breathing (DB). The vertical lines correspond to $\pm\sigma(f_{resp},\epsilon)$, where the local RMSE $\sigma(f_{resp},\epsilon)$ is taken from Figure \ref{fig:6}D. \textbf{(B)} Probabilities of $a_{resp}$ being higher in DB than in NB (yellow, arrow-up), similar in DB and NB (red, tilde), and lower in DB than in NB (blue, arrow-down).}
    \label{fig:7}
\end{figure}

The CNN $\mathcal{C}_{RR,Resp,AFR}$ was used to obtain $\hat{a}_{resp}$ from the clinical data in $\boldsymbol{\mathcal{X}}=[\mathcal{X}_{RR}^{Clin}; \mathcal{X}_{Resp}^{Clin}; \mathcal{X}_{AFR}^{Clin}]$.
The resulting $\hat{a}_{resp}$ for 1-minute segments during normal breathing and deep breathing are shown in Figure \ref{fig:7}.
There was high inter-patient variability in $\hat{a}_{resp}$ in the study population and no clear relation was found between $\hat{a}_{resp}$ during normal breathing and deep breathing.
No significant correlation was found between a change in respiration frequency $\hat{f}_{resp}^{DB}-\hat{f}_{resp}^{NB}$ and a change in respiratory modulation $\hat{a}_{resp}^{DB}-\hat{a}_{resp}^{NB}$.

The vertical lines around $\hat{a}_{resp}$ in Fig. \ref{fig:7}A correspond to $\pm\sigma(f_{resp},\epsilon)$, described in Sec. \ref{sec:predictionmodulationsimdata} and is used for the Monte Carlo sampling described in Sec. \ref{sec:predictionmodulationclindata}.
For 20 subjects, $\hat{a}_{resp}$ was available for at least one segment during normal breathing and one segment during deep breathing (cf. exclusion criteria in Sec. \ref{sec:PostprocessingRRseriesEDRsignals}).
For those 20 subjects, Monte Carlo sampling was used to investigate whether $\hat{a}_{resp}$ is larger during deep breathing than during normal breathing as described in Sec. \ref{sec:predictionmodulationclindata}.
As illustrated in Figure \ref{fig:7}B: it was most likely for 5 patients that the highest $a_{resp}$ was achieved for deep breathing; it was most likely for 5 patients that the lowest $a_{resp}$ was achieved for deep breathing; and it was most likely for 10 patients that neither the highest nor lowest $a_{resp}$ corresponded to deep breathing.

\section{Discussion}
The aim of this study was to develop and evaluate a method to extract respiratory modulation in the AV node conduction properties from ECG data during AF.
To achieve this we extended our AV node model (\cite{Plappert2022}) to account for respiratory modulation by multiplying a time-varying scaling factor to the AV nodal refractory period and conduction delay.
We trained a 1D-CNN on simulated \mbox{1-minute} segments of RR series, respiration signals, and mean arrival rate of atrial impulses which replicate clinical data to predict $a_{resp}$ which quantifies the peak-to-peak amplitude of respiratory modulation.
We evaluated the network on simulated data and the results indicated that $a_{resp}$ can be estimated with an RMSE of 0.066, corresponding to a sixth of the expected range for $a_{resp}$ between 0 and 0.4.

The clinical predictions of $\hat{a}_{resp}$ in Fig. \ref{fig:7} obtained using the trained CNN show a large interpatient variability in $\hat{a}_{resp}$ during deep breathing as well as during normal breathing.
If the magnitude of respiratory modulation in AF would be respiration frequency dependent as it is in NSR (\cite{Angelone1964, Bernardi2001, Russo2017}), then we would expect to see a clear increase in $\hat{a}_{resp}$ during deep breathing.
Results from the Monte Carlo sampling showed that $\hat{a}_{resp}$ increased in response to deep breathing in 5 patients, decreased in 5 patients, and remained the same in 10 patients.
This interpatient variability cannot be explained by differences in respiration rates, since there was no correlation between a change in respiration frequency and a change in respiratory modulation.

In our previous model formulation, we accounted for the ANS-induced changes by introducing constant scaling factors for the refractory period and conduction delay (\cite{Plappert2022}).
With the scaling of AV nodal conduction properties, it was shown that the incorporation of ANS-induced changes in the model allowed better replication of several statistical properties of clinical RR series obtained from tilt tests.
In the present study, this approach was further developed by using a time-varying scaling factor $A^{P}(t)$ to account for respiratory modulation in AV nodal conduction properties based on the assumption that some degree of respiratory modulation generally influences RR series characteristics during AF.
We model respiratory modulation as a joint increase in AV nodal refractoriness and conduction delay in response to exhalation and a joint decrease in AV nodal refractoriness and conduction delay in response to inhalation.
It is known that respiration modulates the parasympathetic activity, where inspiration decreases vagal activity and expiration increases vagal activity (\cite{Katona1970, Russo2017}).
Many electrophysiological (EP) studies have demonstrated that an increase in parasympathetic activity causes an increase in AV nodal conduction delay; studies in dogs reported an increased conduction delay with vagal stimulation (\cite{Irisawa1971, Spear1973, Martin1975, Nayebpour1990a, Pirola1990, Goldberger1999}) and acetylcholine administration (\cite{Priola1983}).
Furthermore, an increase in parasympathetic activity with vagal stimulation in dogs has been demonstrated to increase the AV nodal refractory period (\cite{Spear1973, Nayebpour1990a, Goldberger1999}).
For a decrease in parasympathetic activity with atropine, EP studies demonstrate that the AV nodal conduction delay decreases in dogs (\cite{Irisawa1971}) and humans (\cite{Lister1965, Akhtar1974}), and the AV nodal refractory period also decreases in humans (\cite{Akhtar1974}).

The assumption that some degree of respiratory modulation generally influences the RR series characteristics during AF is also indicated by the fact that some AF patients display clear fluctuations in their RR series matching their respiration frequency (\cite{Rawles1989, Chandler1994, Nagayoshi1997}).
Such fluctuations could also be seen in simulated RR series for some AV node model parameter sets.
During model development, we noticed that an increase in $a_{resp}$ leads to an increase in the relative contribution of the respiration frequency in the frequency spectrum of the RR series with zero-mean $F_{RR}(f_{resp})/\sum_fF_{RR}(f)$ and an increase in the sample entropy of the RR series.
We also noticed that an increase in $f_{resp}$ leads to a decrease in $F_{RR}(f_{resp})/\sum_fF_{RR}(f)$ and an increase in the sample entropy of the RR series.
When averaging over several realizations of RR series (data not shown), $F_{RR}(f_{resp})/\sum_fF_{RR}(f)$ could be clearly seen for most of the parameter sets but is usually masked in individual RR series segments by the irregularity of the RR series.
Using cross-spectral analysis, no simple linear relationship has been found between respiration signal and RR series in AF patients, but a linear relationship was shown in NSR (\cite{Pitzalis1999}).
A possible reason for this is that the relationship between the RR series and respiratory modulation in AV nodal conduction properties during AF is complex and non-linear, emphasizing the need for a model-based approach.
Besides some indications of fluctuations in the RR series, for most of the patients reported in (\cite{Rawles1989, Chandler1994, Nagayoshi1997, Pitzalis1999, Pacchia2011}) and also for the clinical data used in this study, no fluctuations in the RR series matching their respiration frequency were found. 
To match $F_{RR}(f_{resp})/\sum_fF_{RR}(f)$ in the clinical data which was always below 7\%, parameter sets with a higher relative peak spectral energy were excluded from the simulated data (criterion 5 in Sec. \ref{sec:modelAVcond}).
The RR series characteristics of the simulated data differed significantly from both the normal breathing and deep breathing data (cf. Table \ref{tab:3}).
Simulated data with RR series characteristics more similar to the clinical data could be generated by imposing stricter exclusion criteria, e.g., increasing the lower bounds for irregularity and variability set by criteria 3 and 4 in Sec. \ref{sec:modelAVcond}.
However, the simulated data still included signals resembling the clinical data, and the wider range of characteristics likely improved the CNN training by facilitating generalization across a broader range of RR-series.
Nevertheless, it is assumed that by the sheer size of the simulated datasets and the conservative model parameter ranges, there will be simulated RR series in the dataset that resemble the clinical data.

The lead-specific respiration signals were computed using the slope range method which was designed for ECG data during AF (\cite{Kontaxis2020}) and found to be one of the best performing and simplest methods for lead-specific respiration signal extraction (\cite{Varon2020}).
The result of the lead-specific respiration signal extraction can be improved when combining respiration signals from multiple ECG leads with a joint-lead respiration signal.
Previously, the principal component analysis (PCA) has been used to extract joint-lead respiration signals from the clinical data used in this study (\cite{Abdollahpur2022}).
However, the principal components were sensitive to high variance noise as the PCA is based on second-order statistics.
To address this issue, we developed a novel approach for robust fusion of lead-specific respiration signals based on the $\pi$CA (\cite{Sameni2008}).
Under the assumption that the respiration signal has a periodic structure where the respiration frequency and volume between breaths are constant, the $\pi$CA is more suitable for the extraction of joint-lead respiration signals compared to other blind-source separation methods, such as the PCA and basic independent component analysis (ICA).
This is because the $\pi$CA finds the linear mixture of lead-specific respiration signals with maximal periodic structure, whereas the PCA and basic ICA are based on second-order and fourth-order statistics, respectively.
We assume that the respiration frequency and volume between breaths do not vary much in 1-minute segments, making the $\pi$CA a suitable approach for the extraction of short joint-lead respiration signals.
However, considering that the CNN $\mathcal{C}_{RR,Resp,AFR}^{2.5min}$ performs better when using 2.5-minute segments instead of 1-minute segments, another method may be required for the extraction of longer joint-lead respiration signals.

The comparison between the CNN $\mathcal{C}_{RR,Resp,AFR}$ and the linear regression shown in Fig. \ref{fig:5} demonstrate that the relation between the RR series, respiration signal and mean atrial arrival rate to $a_{resp}$ is not simple and not linear.
In this study, we only investigate the performance of one basic CNN architecture.
While some variations on this were tested during the neural network development, no extensive investigation has been performed and there may be better neural network architectures for this task.
The CNN requires the RR series for the prediction of $a_{resp}$ and the mean atrial arrival rate always improved the prediction.
In this evaluation, however, $\mu$ was set to the correct value; we did not account for estimation errors that are most likely present in real data since AFR provides a crude estimate of the atrial arrival rate.
Moreover, the addition of the respiration signal only improves the prediction when of sufficient quality as quantified by $\epsilon$.
The linear dependence between $\eta$ and $\epsilon$ support our assumption of $\epsilon$ as a marker of respiration signal quality (cf. Sec. \ref{sec:simRRRespSignals}).
Whereas the addition of the respiration signal and mean atrial arrival rate can improve the prediction of $\hat{a}_{resp}$, a method based on RR series only is less sensitive to noise in the recordings.
Potentially, the RR series could be extracted from pulse watch data, allowing for continuous monitoring of $a_{resp}$ in a wide range of applications.

The performance of the CNN is dependent on $f_{resp}$ and $\epsilon$ (cf. Fig. \ref{fig:6}), where $f_{resp}$ appears to have a larger impact on the performance than $\epsilon$.
The marker of respiration signal quality $\epsilon$ was not used as an exclusion criterion for 1-minute segments, because the addition of $\mathcal{X}_{Resp}^{Sim}$ to the input only slightly improved the accuracy of the $\hat{a}_{resp}$ prediction and the influence of $\epsilon$ on the RMSE compared to $f_{resp}$ was small.
Instead, $\epsilon$ was used to choose the best combination of non-overlapping 1-minute segments.
Interestingly, the performance of the CNNs $\mathcal{C}_{RR}$, $\mathcal{C}_{AFR}$, $\mathcal{C}_{RR,AFR}$ still show a slight dependence on $\epsilon$ although this parameter quantifies the non-periodicity and signal quality of $\mathcal{X}_{Resp}^{Sim}$ (cf. Fig. \ref{fig:6}).
This suggests that $\epsilon$ carries information about the RR interval series, and may indicate that the distribution of AV node model parameters varies over different $\epsilon$ and that different subsets of model parameters result in different local RMSEs.
One possible explanation why the impact of $f_{resp}$ on the performance is prominent may be that there are fewer respiratory cycles in the 1-minute segment at lower $f_{resp}$.
When using 2.5-minute segments in the input data, the performance of the CNN $\mathcal{C}_{RR,Resp,AFR}^{2.5min}$ improved overall, especially at lower $f_{resp}$.
The segment length was set to 1 minute in this study due to the recording length of 1 minute during deep breathing. 

There are several limitations of the present study.
We assume for simplicity that the variations in AV nodal refractoriness are similar to the variations in AV nodal conduction delay.
We also assume that the variations in AV nodal refractoriness and conduction delay are similar between SP and FP.
Moreover, the model does not include phase shifts between the RR series and respiration signal for different respiration frequencies (\cite{Angelone1964}), or effects of respiration volume (\cite{Grossman2007}).
Hence, a different scaling for the refractory period and conduction delay, a different scaling for the SP and FP, a phase shift between the RR series and respiration signal, and an inclusion of respiration volume might form interesting directions for future model improvements.
We did not account for respiratory modulation in the AA series, because the modulation is small during AF (\cite{Abdollahpur2022,Celotto2020}).
When choosing the bounded uniform distribution of $a_{resp}$ for the training and testing dataset, we made a tradeoff between bias and variance.
The reason why $a_{resp}$ was randomly drawn from ${\displaystyle {\mathcal {U}}{[-0.1,0.5]}}$ in the training data and randomly drawn from ${\displaystyle {\mathcal {U}}{[0,0.4]}}$ in the testing data of the CNN is to reduce the bias in the $\hat{a}_{resp}$ prediction (cf. Fig. \ref{fig:5}).
Without extending the range of $a_{resp}$ in the training data, the sample mean of the $\hat{a}_{resp}$ diverged more from $a_{resp}$ at values close to $0$ and $0.4$.
However, the accuracy of the CNNs decreased by extending the range of $a_{resp}$ in the training data.
While plenty of simulated data with $a_{resp}$ ground truth can be generated using the AV node model, there was no $a_{resp}$ ground truth available for the clinical dataset and its size was limited.

\section{Conclusion}
We presented an extended AV node model that accounts for respiratory modulation in conduction delay and refractory period.
We trained a 1D-CNN to predict the respiratory modulation in the AV node with simulated RR series, respiration signal, and mean atrial arrival rate which replicate clinical ECG-derived data.
Using simulated data, we demonstrated that the respiratory modulation can be predicted using the 1D-CNN from RR series alone and that the prediction is improved when adding a respiration signal and AFR.
Results from analysis of clinical ECG data of 20 patients with sufficient signal quality suggest that the respiratory modulation decreased in response to deep breathing for five patients, increased for five patients, and remained similar for ten patients, indicating a large inter-patient variability.

\section*{Conflict of Interest Statement}
The authors declare that the research was conducted in the absence of any commercial or financial relationships that could be construed as a potential conflict of interest.

\section*{Author Contributions}

FP, MW, PP, and FS contributed to the conception and design of the study.
GE was responsible for the deep breathing test experiment. 
FP performed the ECG processing, developed the joint-lead EDR signal extraction, designed and developed the extended AV node model and convolutional neural network, wrote the original draft of the manuscript, and produced the figures, with supervision from MW and FS.
MW, PP, and FS contributed to the funding acquisition.
PP and GE contributed to the clinical interpretation of the results.
All authors contributed to the manuscript revision, read, and approved the submitted version.

\section*{Funding}
The research was supported by the Swedish Research Council
(grant VR 2019–04272), the Crafoord Foundation (grant
20200605), and the Swedish Heart-Lung foundation (no 2020-0173).

\section*{Acknowledgments}
The computations were enabled by resources provided by the National Academic Infrastructure for Supercomputing in Sweden (NAISS) and the Swedish National Infrastructure for Computing (SNIC) at Lund University partially funded by the Swedish Research Council through grant agreements no. 2022-06725 and no. 2018-05973.
The Swedish Heart and Lung foundation was the main funding body of the SCAPIS cohort. 
SCAPIS was also supported by grants from the Knut and Alice Wallenberg Foundation, the Swedish Research Council, and Sweden’s Innovation agency. 

\section*{Supplemental Data}
The Supplementary Material for this article can be found online at: \href{https://github.com/PlappertF/ECG-based_estimation_of_respiratory_modulation_of_AV_nodal_conduction_during_atrial_fibrillation}{https://github.com/PlappertF/ECG-based\_estimation\_of\_respiratory\_modulation\_of\_AV\_nodal\_conduction\_during\_atrial\_fibrillation}

\section*{Data Availability Statement}
The data analyzed in this study is subject to the following licenses/restrictions: Requests to access data from the SCAPIS cohort should be directed to info@scapis.org (www.scapis.org/data-access/).
The code for the AV node model together with a user example can be found in the Supplementary Material.

\bibliographystyle{unsrtnat}
\bibliography{manuscript}

\end{document}